\documentclass[twocolumn]{aastex63}
\let\tablenum\relax

\usepackage{amsmath,bm}
\usepackage{multirow}
\usepackage{hyperref}
\usepackage{nameref}
\usepackage{siunitx}

\received{XXX}
\revised{YYY}
\accepted{ZZZ}

\submitjournal{ApJ}

\shorttitle{environments and clustering of JWST AGNs}
\shortauthors{Lin et al.}

\graphicspath{{./}{figures/}}
 
\newcommand{\ha}{H$\alpha$}

\newcommand{\MBH}{$M_{\rm BH}$}

\newcommand{\xj}[1]{{{#1}}}
\newcommand{\NAGN}{28}
\newcommand{\NHAE}{782}

\begin{document}

\title{The Large-scale Environments of Low-luminosity AGNs at $3.9 < z < 6$ and Implications for Their Host Dark Matter Halos from a Complete NIRCam Grism Redshift Survey}

\author[0000-0001-6052-4234]{Xiaojing Lin}
\affiliation{Department of Astronomy, Tsinghua University, Beijing 100084, China}
\email{xiaojinglin.astro@gmail.com}
\affil{Steward Observatory, University of Arizona, 933 N Cherry Ave, Tucson, AZ 85721, USA}

\author[0000-0003-3310-0131]{Xiaohui Fan}
\affiliation{Steward Observatory, University of Arizona, 933 N Cherry Ave, Tucson, AZ 85721, USA}

\author[0000-0002-4622-6617]{Fengwu Sun}
\affiliation{Center for Astrophysics $|$ Harvard \& Smithsonian, 60 Garden St., Cambridge, MA 02138, USA}

\author[0000-0002-1574-2045]{Junyu Zhang}
\affiliation{Steward Observatory, University of Arizona, 933 N Cherry Ave, Tucson, AZ 85721, USA}

\author[0000-0003-1344-9475]{Eiichi Egami}
\affiliation{Steward Observatory, University of Arizona, 933 N Cherry Ave, Tucson, AZ 85721, USA}

\author[0000-0003-4337-6211]{Jakob M.\ Helton}
\affiliation{Steward Observatory, University of Arizona, 933 N Cherry Ave, Tucson, AZ 85721, USA}

\author[0000-0002-7633-431X]{Feige Wang}
\affiliation{Department of Astronomy, University of Michigan, 1085 South University Avenue, Ann Arbor, MI 48109, USA}

\author[0000-0002-4321-3538]{Haowen Zhang}
\affiliation{Steward Observatory, University of Arizona, 933 N Cherry Ave, Tucson, AZ 85721, USA}

\author[0000-0002-8651-9879]{Andrew J.\ Bunker }
\affiliation{Department of Physics, University of Oxford, Denys Wilkinson Building, Keble Road, Oxford OX13RH, UK}

\author[0000-0001-8467-6478]{Zheng Cai}
\affiliation{Department of Astronomy, Tsinghua University, Beijing 100084, China}

\author[0000-0001-7673-2257]{Zhiyuan Ji}
\affiliation{Steward Observatory, University of Arizona, 933 N Cherry Ave, Tucson, AZ 85721, USA}

\author[0000-0002-5768-738X]{Xiangyu Jin}
\affiliation{Steward Observatory, University of Arizona, 933 N Cherry Ave, Tucson, AZ 85721, USA}

\author[0000-0002-4985-3819]{Roberto Maiolino}
\affiliation{Kavli Institute for Cosmology, University of Cambridge, Madingley Road, Cambridge, CB3 0HA, UK }
\affiliation{Cavendish Laboratory - Astrophysics Group, University of Cambridge, 19 JJ Thomson Avenue, Cambridge, CB3 0HE, UK } 
\affiliation{Department of Physics and Astronomy, University College London, Gower Street, London WC1E 6BT, UK}

\author[0000-0003-4924-5941]{Maria Anne Pudoka}
\affiliation{Steward Observatory, University of Arizona, 933 N Cherry Ave, Tucson, AZ 85721, USA}

\author[0000-0002-5104-8245]{Pierluigi Rinaldi}
\affiliation{Steward Observatory, University of Arizona, 933 N Cherry Ave, Tucson, AZ 85721, USA}

\author[0000-0002-4271-0364]{Brant Robertson}\affiliation{Department of Astronomy and Astrophysics, University of California, Santa Cruz, 1156 High Street, Santa Cruz, CA 95064, USA}

\author[0000-0002-8224-4505]{Sandro Tacchella}
\affiliation{Kavli Institute for Cosmology, University of Cambridge, Madingley Road, Cambridge, CB3 0HA, UK
}
\affiliation{Cavendish Laboratory, University of Cambridge, 19 JJ Thomson Avenue, Cambridge, CB3 0HE, UK}

\author[0000-0003-0747-1780]{Wei Leong Tee
}
\affiliation{Steward Observatory, University of Arizona, 933 N Cherry Ave, Tucson, AZ 85721, USA}

\author[0000-0001-6561-9443]{Yang Sun}
\affiliation{Steward Observatory, University of Arizona, 933 N Cherry Ave, Tucson, AZ 85721, USA}

\author[0000-0001-9262-9997]{Christopher N.\ A.\ Willmer}
\affiliation{Steward Observatory, University of Arizona, 933 N Cherry Ave, Tucson, AZ 85721, USA}

\author[0000-0002-4201-7367]{Chris Willott}
\affiliation{NRC Herzberg, 5071 West Saanich Rd, Victoria, BC V9E 2E7, Canada}

\author[0000-0003-3307-7525]{Yongda Zhu}
\affiliation{Steward Observatory, University of Arizona, 933 N Cherry Ave, Tucson, AZ 85721, USA}



\correspondingauthor{Xiaojing Lin}

\begin{abstract}
We study the large-scale environments and clustering properties of 28 low-luminosity AGNs at $z=3.9-6$ in the GOODS-N field. Our sample, identified from the JWST NIRCam Imaging and WFSS data in CONGRESS and FRESCO surveys with either broad \ha\ emission lines or V-shaped continua, are compared to \NHAE\ \ha\ emitters (HAEs) selected from the same data. These AGNs are located in diverse large-scale environments and do not preferentially reside in denser environments compared to HAEs. Their overdensity field, $\delta$, averaged over (15 $h^{-1}$cMpc)$^3$, ranges from $-0.56$  to 10.56, and shows no clear correlation with broad-line luminosity, black hole (BH) masses, or the AGN fraction. It suggests that $> 10$ cMpc structures do not significantly influence BH growth. We measure the two-point cross-correlation function of AGNs with HAEs, finding a comparable amplitude to that of the HAE auto-correlation. This indicates similar bias parameters and host dark matter halo masses for AGNs and HAEs. The correlation length of field AGNs is 4.26 $h^{-1}$cMpc, and 7.66 $h^{-1}$cMpc at $3.9 < z < 5$ and $5 < z < 6$, respectively. We infer a median host dark matter halo mass of $\log (M_h/M_\odot)\approx 11.0-11.2$ and host stellar masses of $\log (M_\star/M_\odot) \approx 8.4-8.6$ by comparing with the \textsc{UniverseMachine} simulation. Our clustering analysis suggests that low-luminosity AGNs at high redshift reside in normal star-forming galaxies with overmassive BHs. They represent an intrinsically distinct population from luminous quasars and could be a common phase in galaxy evolution.
\end{abstract}


\section{Introduction}

Over the past few decades, the discovery of UV-luminous quasars at $z \gtrsim 6$ has reshaped and challenged our understanding of supermassive black holes (SMBHs) in the early Universe \citep[e.g.,][]{Banados2018, Yang2020, Wang2021, Fan2023}. 
These $z > 6$ quasars, hosting SMBHs exceeding $10^9\,M_\odot$ \citep[e.g.,][]{Yang2021, Farina2022}, have motivated extensive work on models of BH growth modes and seeding mechanisms \citep[e.g.,][]{Inayoshi2022, Regan2024, Cammelli2024}. Multiple approaches have been proposed to constrain quasar lifetimes and duty cycles, including studies of quasars' proximity zones \citep{Eilers2017, Eilers2020, Davies2020}, damping wing features \citep{Davies2019, Hennawi2024}, and clustering properties \citep{Eilers2024, Pizzati2024a}. These independent approaches suggest quasar lifetimes of $\sim 10^6$ yr, with low duty cycles ($\ll 1$).  These findings imply a very short timescale for quasar accretion and extremely rapid BH growth with low radiative efficiency \citep[$\lesssim 0.1\%$, ][]{Davies2019, Eilers2024}. Alternatively, the bulk of SMBH growth history might have been enshrouded by dust and thus missed by current surveys \citep{Comastri2015, Ni2020, Lambrides2024a}.  The demographics of a broader BH population is the key to understanding the early BH assembly history.

The unprecedented near-infrared capabilities of the James Webb Space Telescope \citep[JWST,][]{Gardner2023} have brought new insights. Since its launch, JWST has revealed new populations of low-luminosity active galactic nuclei (AGNs) at $z > 4$, many of which are nicknamed as ``little red dots'' due to their compact morphology and unique spectral energy distributions (SEDs) in the near-IR wavelengths \citep[e.g.,][]{Matthee2024, Greene2024, Kokorev2024, Lin2024, Akins2024, Perez-Gonzalez2024, Rinaldi2024}. They often exhibit broad Balmer emission lines with full width at half maximum (FWHM) exceeding 1000\,km\,s$^{-1}$. Assuming these broad emission lines originate from the AGN broad-line region, they suggest SMBHs with masses ranging from $10^6 M_\odot$ to $\lesssim 10^9 M_\odot$ \citep[e.g., ][]{Greene2024, Lin2024}. While JWST has revealed these previously unseen objects, their nature and connection to UV-luminous quasars remain elusive. For instance, there is an ongoing debate regarding the origin of the so-called ``V-shaped'' continua observed in many of these objects, which exhibit a reddened rest-frame optical continuum and a bluer UV continuum slope \citep[e.g.,][]{Ma2024, Li2024}. It is also difficult to explain the absence of a hot dust torus \citep{Casey2024, Perez-Gonzalez2024, Williams2024}, the weakness of X-rays \citep{Yue2024b, Maiolino2024}, and the lack of variability in both UV and optical bands \citep{Tee2024, Zhang2024, Kokubo2024}, which are typical features of UV-luminous quasars. The high spatial resolution of NIRCam short-wavelength imaging reveals that $\sim 30\%$ of LRDs exhibit signs of interaction, suggesting a potential link to merger-driven activities \citep{Rinaldi2024}.  Some theoretical hypotheses propose that the broad emission lines may not originate from SMBHs, but rather from unusual galaxy kinematics or outflows \citep{Baggen2024, Kokubo2024}. In this paper, we refer to these objects (broad Balmer line emitters or ``little red dots'') as low-luminosity AGNs, as they literally exhibit the defining characteristics of AGNs---active and compact. While we assume the broad lines originate from BHs,  as is commonly done in other works, we note the possibility of their non-BH origins. The measurements presented in this paper, and comparisons with those of their co-eval galaxy populations, are independent of the assumptions of the origins of their energy source in most cases. 

To understand the nature of low-luminosity AGNs, it is crucial to constrain their large-scale environments and host dark matter halos within the context of cosmological structure formation models.  
Currently, our knowledge about the host galaxies of low-luminosity AGNs largely comes from SED modeling. However, this method is limited by assumptions about the intrinsic characteristics, leading to significant degeneracy \citep{Ma2024, Leung2024, Akins2024}. It can result in a wide range of stellar masses, from extremely massive hosts ($\sim 10^{11} M_\odot$) to typical star-forming galaxies ($\sim 10^8 M_\odot$), depending on how AGN and galaxy light contributions are tuned \citep[e.g.,][]{Wang2024}. The clustering of low-luminosity AGNs can break this degeneracy by directly comparing their large-scale environments and dark matter halo masses with those of galaxies of similar stellar masses.  Clustering analysis also provides insights to bridge the gap between low-luminosity AGNs and UV-bright quasars. 
UV luminous quasars are characterized by strong large-scale clustering over a wide redshift range, consistent with host halo masses of $\sim 10^{12.5} M_\odot$ at $z>6$ \citep[e.g.,][]{Chen2022, Arita2023, Eilers2024, Pudoka2024}.
If low-luminosity AGNs represent the dust-obscured phase of UV-luminous quasars \citep[e.g.,][]{Lyu2024}, they should inhabit similar massive dark matter halos and share comparable clustering properties. Observational studies of AGN environments and clustering properties can also inform theoretical models. For example, super-Eddington accretion has been proposed to explain some characteristics of low-luminosity AGNs, such as the V-shaped SEDs, X-ray weaknesses, and lack of variability \citep{Lambrides2024, Inayoshi2024b, Kokubo2024, Trinca2024}. Super-Eddington accretion is predicted to occur only in low-mass hosts ($\sim 10^{10}\,M_\odot$) where the AGN duty cycle is low \citep[$<1\%$, ][]{Pizzati2024}. Recently, \cite{Schindler2024} reported a low-luminosity AGN in an overdensity, with a minimum host halo mass of $10^{12.3}\,M_\odot$. However, systematic studies involving a larger sample of AGNs are still required. \cite{Matthee2024b} examined the environments within 1\,cMpc of six AGNs in the Abell 2744 lensing cluster field. Nonetheless, non-linear physical processes may dominate on such small scales \citep{Harikane2016, Herard_Demanche_2023}. Their analysis does not extend to larger scales due to survey volume limitations. A clustering analysis on scales $\gtrsim$10\, comoving Mpc (cMpc) is crucial to probe the linear two-halo term and achieve more accurate constraints on halo masses \citep[e.g.,][]{Arita2024}.

In this context, the JWST/NIRCam Wide Field Slitless Spectroscopy \citep[WFSS,][]{Greene2016, Rieke2023} provides a unique and powerful tool for clustering analyses of low-luminosity AGNs. The JWST/NIRCam WFSS has demonstrated great efficiency in studying high-redshift emission-line galaxies. It is highly effective in identifying broad-line AGNs because of its relatively high spectral resolution, and in probing large-scale structures because of its large field-of-view \citep[FoV; e.g.,][]{Sun2022a, Lin2024, Helton2024}.  
The selection function is simpler to model compared to the spectroscopic observations with pre-selection (e.g., through JWST NIRSpec). In this work, we compile AGNs and galaxies from a Complete NIRCam Grism Redshift Survey (CONGRESS) in the GOODS-N field. The dataset includes grism observations from the Cycle-1 program FRESCO \citep[``First Reionization Epoch Spectroscopically Complete Observations", GO-1895, PI: Oesch,][]{Oesch2023} and the Cycle-2 program CONGRESS (GO-3577, PI Eiichi \& Sun, Sun et al. in prep). The two programs target the same area using the F444W and F356W grisms, respectively, together covering a total wavelength range of $3.1-5.0$\,\micron. We collect a large number of low-luminosity AGNs at $4\lesssim z<6$ through their broad \ha\ emission lines and broadband photometric properties, and simultaneously map their surrounding environments with \ha\ emitters at the same redshift.

The paper is organized as follows. In \S\ref{sec:data_sample}, we introduce the datasets and methods to select galaxies and AGNs. In \S\ref{sec:DiverseEnv}, we explore the large-scale environments of low-luminosity AGNs. A clustering analysis based on two-point correlation functions is presented in \S\ref{sec:clustering}. Finally, in \S\ref{sec:um}, we discuss the potential implications of our measurements for AGN evolution. Throughout this work, a flat $\Lambda$CDM cosmology is assumed, with $\rm H_0 = 70~km~s^{-1}~Mpc^{-1} $, $\Omega_{\Lambda,0} = 0.7$ and  $\Omega_{m,0}=0.3$.  We define $h={\rm H_0}/100=0.7$.

\bigskip

\section{Data and Sample}\label{sec:data_sample}

\subsection{Imaging and photometric catalog}
We use the images and photometric catalog in the GOODS-N field from the JADES Data Release 3\footnote{\url{https://archive.stsci.edu/hlsp/jades}} \citep{JADES_DR3}.  The JADES JWST/NIRCam images in the GOODS-N field include observations from GTO programs 1181 (PI Eisenstein) and GO programs 1895 (FRESCO, PI Oesch), spanning the F090W, F115W, F150W, F182M, F200W, F210M, F277W, F335M, F356W, F410M, and F444W filters. The final mosaics cover from 56 arcmin$^{2}$ in F090W to 83 arcmin$^{2}$ in F444W \citep{JADES_DR3}.  We refer to \cite{JADES_Eisenstein2023} for a detailed description of the JADES survey design, and \cite{Rieke2023} and Robertson et al. (in prep.) for the imaging data reduction. For the direct images used for the grism spectra, the JADES images achieve a 5$\sigma$ point-source depth of 29.38 mag in F444W and 29.97 mag in F356W,  with aperture-corrected photometry using an $r=0\farcs15$ circular aperture.

The JADES GOODS-N photometric catalog includes multi-band photometry in the 11 JWST/NIRCam filters as mentioned above, and five HST/ACS filters (F435W, F606W, F775W, F814W, and F850LP). The HST/ACS photometry is based on images from the Hubble Legacy Fields project  \citep[HLF,][]{Illingworth2017}. We refer to \cite{Robertson2024} for detailed source detection and photometry measurement methods.  The photometric redshifts are estimated utilizing the $r=0\farcs1$ circular aperture photometry, using \texttt{EAZY} \citep{Brammer2008} with galaxy spectral energy distribution (SED) templates optimized for high-redshift sources \citep{Hainline2024}.

\subsection{JWST/NIRCam Grism spectroscopy}
JWST/NIRCam WFSS observations of the GOODS-N field were obtained in both F356W and F444W filters.   The Cycle-1 program, FRESCO, covers 62 arcmin$^{2}$ of the GOODS-N field through the F444W filter and row-direction grism (Grism R). The FRESCO observations in GOODS-N were split into eight pointings, and the exposure time is 8$\times$880\,s per pointing. The Cycle-2 program, CONGRESS,  targets the same areas in the GOODS-N field observed by FRESCO.
CONGRESS adopts the F356W filter and Grism R.
CONGRESS includes 12 pointings, and the exposure time is 8$\times$472\,s per pointing.  The combination of F356W and F444W grism observations results in a total wavelength coverage at 3.1--5.0\,\micron. The overlap area between JADES, FRESCO, and CONGRESS is about 62 arcmin$^2$, as illustrated in the bottom
panel of Figure \ref{fig:distribution}.

The grism data reduction and spectral extraction are detailed in \cite{Sun2025}. Here we briefly summarize the procedures. For individual exposures of grism data and their corresponding short-wavelength (SW) direct images, we performed flat-fielding, subtracted a sigma-clipped median sky background, and aligned the World Coordinate System (WCS) frames.  
We then measured the astrometric offsets between the direct images and the JADES DR3 GOODS-N catalog. The offsets were added to the 
spectral tracing model for accurate wavelength calibration.  The spectral tracing and wavelength calibration are based on the Commissioning, Cycle-1, and Cycle-2 calibration data taken in the SMP-LMC-58 field up to June 2024 (PID 1076, 1479, 1480, and 4449; see \citealt{Sun2023})\footnote{\url{https://github.com/fengwusun/nircam_grism}}.
The flux calibration is based on Cycle-1 calibration data (PID: 1076, 1536, 1537, 1538). 
We optimally extracted the 1D spectra based on the source morphology \citep{Horne1986}.

\subsection{\ha\ emitters  at $3.9<z<6$ }\label{sec:hae_sample}
We refer readers to \citet{Lin_GDN_HAE} for detailed descriptions of the selection procedure for line emitters. In brief, we use 51-pixel median filtering to remove the continua, extract line-only spectra, and detect emission lines in both 1D and 2D spectra. We have developed a semi-automated algorithm to identify emission-line galaxies across the redshift range $0 < z < 9$ using the photometric redshifts in the JADES catalog as priors. As a result, we selected 936 HAEs across $3.75 < z < 6.6$ with F356W and F444W Grism R in the overlapping regions of the CONGRESS and JADES imaging footprint. This work focuses on a subsample of these HAEs within the redshift range of $3.9 < z < 6$.  We measured the line flux by fitting the emission line with Gaussian models convolved with the NIRCam grism line spread functions as calibrated by \cite{Sun2025}. The emitter catalog and line flux measurements will be publicly released along with \cite{Sun2025}.

In this work, we only consider HAEs and AGNs with \ha\ luminosities greater than $L_{\rm H\alpha} > 10^{41.5}~{\rm erg~s^{-1}}$.  This is because the clustering analysis relies on luminosity function (LF) measurements, and the completeness correction for lower luminosity bins in the \ha\ LFs is affected by large systematic uncertainties. \xj{For AGNs, this cut is applied to the total $L_{\rm H\alpha}$ to account for selection effects in the detectability of the H$\alpha$ line.} We further group galaxies with separations smaller than 10 physical kpc and 500 \si{km\,s^{-1}} into a system, assuming they are interacting and gravitationally bound. This separation corresponds to approximately 1\farcs5\ at $z \approx 4 - 6$, consistent with the definition of a galaxy system applied to $z \sim 6$ [\ion{O}{3}] emitters \citep{Matthee2023, Eilers2024} and well within the defining separation for mergers \citep[e.g.,][]{Puskas2025}.
This differs from the criteria set for the LF calculation in \citet{Lin_GDN_HAE}, but it follows the same method used for the HAE clustering analysis in that paper. We finally obtain \NHAE\ systems at $3.9<z<6$. Throughout this paper, we refer to these combined systems as galaxies for simplicity. 

\subsection{AGN sample at $3.9<z<6$}\label{sec:agn_sample}
We selected AGNs from the parent HAE sample using three different sets of criteria: WFSS-based, V-shaped-based, and NIRSpec-based. All these AGNs fall within the grism footprint. The selection criteria are outlined below.

\begin{itemize}
    \item[(1)] WFSS-selected broad-line AGNs. With the F356W and F444W grism covering 3.1--5.0\,\micron,  \cite{Junyu2024} have identified 19 broad HAE with compact morphology 
    \footnote{The criteria for compactness is defined based on circular aperture photometry $\frac{\rm{F444W\ flux\ within\ r=0\farcs2}}{\rm{F444W\ flux\ within\ r=0\farcs1}}\leq1.2$, as justified in \cite{Junyu2024}.} and FWHMs of broad \ha\ components exceeding 1000\,km s$^{-1}$.  In our sample, seven of these AGNs were previously reported in \cite{Matthee2024}.

    \item[(2)] NIRSpec-selected broad-line AGNs.  \cite{Maiolino2023} identified high-redshift AGNs through prominent broad \ha\ emission lines in the NIRSpec R1000 grating spectra. In the GOODS-N field, they reported five objects at $4<z<6$ with single robust broad components.  We have excluded the tentatively detected broad-line candidates and the dual broad-line candidates from their sample. Four of the five AGNs fall within the WFSS footprint but do not overlap with the WFSS-selected AGNs. These four objects show bright \ha\ emission lines in the grism spectra, although the broad components are overwhelmed by the background noise because of the grism's lower sensitivity and higher background level.

    \item[(3)] V-shaped continuum AGNs. We have searched for compact HAE with V-shaped spectral energy distributions (SEDs) among the grism-selected HAEs. V-shaped SEDs, characterized by blue UV continuum slopes ($\beta_{\rm UV}$) and red optical continuum slopes ($\beta_{\rm opt}$), have been demonstrated as effective characteristics of high-redshift AGNs \citep[e.g.,][]{Greene2024} especially when the broad \ha\ fluxes are below the grism detection limit. 
    We do not require a broad \ha\ line detection for V-shaped-selected AGNs.

First, we selected \ha\ emitters with a half-light radius $< 0\farcs2$ in either the F444W or F356W filter. The robust spectroscopic redshifts ($z_{\rm spec}$) help us avoid contamination from brown dwarfs. We measured the UV continuum slope $\beta_{\rm UV}$ by fitting a power-law model to the rest-frame range 1350--3000\,\AA\ using the small-Kron-aperture \textsc{KRON\_S} photometry (see JADES NIRCam data release, \citealt{Rieke2023}). The optical continuum slope $\beta_{\rm opt}$ is derived from 4000--8000\,\AA. We excluded bands that overlap with the H$\beta$ + [\ion{O}{3}], \ha\ lines, or the Balmer break around $3645$\,\AA. We require at least two filters spanning over 1000 \AA\ in the restframe to determine $\beta_{\rm opt}$. This strict requirement ensures that the measured slopes represent the pure continuum, free from contamination by strong emission lines and flux uncertainties.

Among the 23 broad-line selected AGNs identified from WFSS and NIRSpec above, seven exhibit significant V-shaped SEDs with $\beta_{\rm UV}<0$ and $\beta_{\rm opt}>0$. Of the remaining 16 AGNs, seven have $\beta_{\rm opt}<0$, while the other nine lack adequate band coverage to reliably determine $\beta_{\rm opt}$.

In addition to the broad-line AGNs above, we identify five more HAEs with robust measurements of $\beta_{\rm UV}<0$ and $\beta_{\rm opt}>0$.

\end{itemize}

\begin{figure}[!t]
    \centering
    \includegraphics[width=\linewidth]{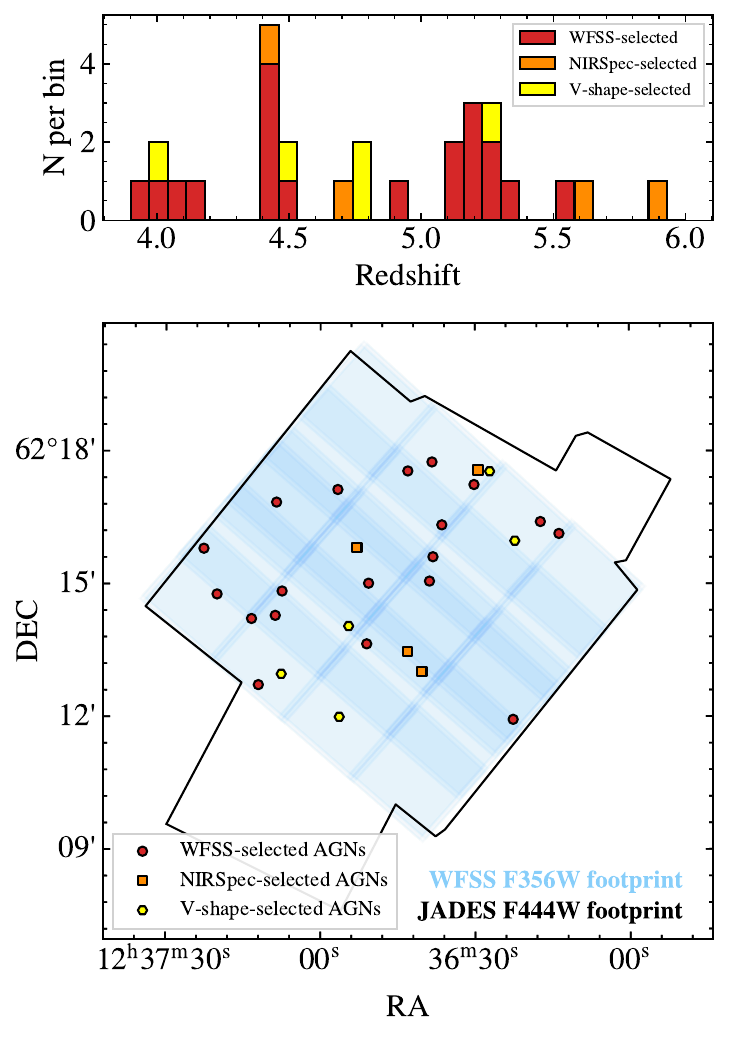}
    \caption{\textit{Top}: the redshift distributions of AGNs in the GOODS-N field. The red, orange, and yellow histograms represent WFSS-selected, NIRSpec-selected, and V-shaped-selected samples, respectively. \textit{Bottom}:  the spatial distribution of AGNs. The blue shaded regions are the F356W WFSS (GO-3577, CONGRESS) footprint, and the black lines enclose the F444W image footprint of JADES  DR3 (including FRESCO GO-1895).  AGNs at $3.9 < z < 6$ are marked as red dots, orange squares, and yellow hexagons for WFSS-selected, NIRSpec-selected, and V-shaped selected sources, respectively.}
    \label{fig:distribution}
\end{figure}

Following the three selection criteria described above, we eventually selected a sample of \NAGN\ AGNs at $z = 3.9 - 6$ within the WFSS footprint for the parent HAE sample. Seven of them are also included in the photometrically selected sample in \cite{Rinaldi2024}. The information of these \NAGN\  AGNs is summarized in Table \ref{tab:broadline_agn} and \ref{tab:vshaped_agn}. Their spatial distribution is shown in Figure \ref{fig:distribution}. All of these AGNs display \ha\ emission lines with $L_{\rm H\alpha} > 10^{41.5} \, \mathrm{erg \, s^{-1}}$ in the grism spectra, satisfying the selection criteria for HAEs in the clustering analysis.

\begin{table*}[!ht]
	\hspace*{-1cm}\begin{tabular}{cccccccccc}
        \hline\hline
			ID & RA & DEC & $z_{\rm spec}$ & selection & $\log L_{\rm H\alpha, broad}$ & FWHM$_{\rm H\alpha, broad}$ & $\log M_{\rm BH}$ & $\beta_{\rm UV}$ & $\beta_{\rm opt}$ \\
			 &  &  &  &  & (erg s$^{-1}$) & (km s$^{-1}$) & ($M_{\odot}$) &  &  \\
             \hline
			1087315 & 189.3336 & 62.2462 & 3.91 & WFSS & 42.09$\pm$0.05 & 1514$\pm$183 & 7.01$\pm$0.11 & -0.44$\pm$1.02 & -1.47$\pm$0.13 \\
			1082263 & 189.2126 & 62.2274 & 3.98 & WFSS & 41.96$\pm$0.07 & 1084$\pm$205 & 6.65$\pm$0.17 & -1.70$\pm$0.09 & -0.20$\pm$0.09 \\
			1089568 & 189.1518 & 62.2722 & 4.05 & WFSS & 42.22$\pm$0.04 & 1462$\pm$143 & 7.04$\pm$0.09 & -1.16$\pm$0.43 & -1.20$\pm$0.10 \\
			1029154 & 189.1590 & 62.2602 & 4.17 & WFSS & 42.33$\pm$0.04 & 2003$\pm$225 & 7.38$\pm$0.10 & -1.56$\pm$0.19 & 0.74$\pm$0.17 \\
			1008411 & 189.2111 & 62.2503 & 4.41 & WFSS & 42.11$\pm$0.11 & 3282$\pm$757 & 7.72$\pm$0.21 & -0.74$\pm$0.32 & 0.72$\pm$0.12 \\
			1008671 & 189.1618 & 62.2511 & 4.41 & WFSS & 42.54$\pm$0.02 & 2273$\pm$95 & 7.59$\pm$0.04 & -1.53$\pm$0.13 & 0.13$\pm$0.11 \\
			1086855 & 189.2865 & 62.2381 & 4.41 & WFSS & 42.40$\pm$0.06 & 1724$\pm$271 & 7.28$\pm$0.14 & -1.01$\pm$0.19 & 0.58$\pm$0.12 \\
			1086784 & 189.3057 & 62.2369 & 4.41 & WFSS & 42.15$\pm$0.07 & 3179$\pm$504 & 7.71$\pm$0.14 & -0.13$\pm$1.26 & -0.10$\pm$0.15 \\
			1033320 & 189.1258 & 62.2874 & 4.48 & WFSS & 42.04$\pm$0.08 & 1951$\pm$398 & 7.22$\pm$0.19 & -1.65$\pm$0.12 & -0.67$\pm$0.26 \\
			1085355 & 189.0944 & 62.1990 & 4.88 & WFSS & 42.35$\pm$0.03 & 1801$\pm$158 & 7.29$\pm$0.08 & -1.76$\pm$0.70 & -- \\
			1090253 & 189.2855 & 62.2808 & 5.09 & WFSS & 42.41$\pm$0.03 & 1455$\pm$105 & 7.13$\pm$0.07 & -1.05$\pm$0.75 & -- \\
			1014406 & 189.0721 & 62.2734 & 5.15 & WFSS & 42.34$\pm$0.07 & 3212$\pm$639 & 7.81$\pm$0.18 & -1.61$\pm$0.16 & -- \\
			1034620 & 189.1598 & 62.2959 & 5.19 & WFSS & 42.68$\pm$0.02 & 1077$\pm$71 & 6.99$\pm$0.06 & -1.43$\pm$0.06 & -- \\
			1090549 & 189.2359 & 62.2855 & 5.19 & WFSS & 42.21$\pm$0.07 & 1722$\pm$307 & 7.19$\pm$0.16 & -1.88$\pm$0.24 & -- \\
			9994014 & 189.3001 & 62.2120 & 5.23 & WFSS & 42.61$\pm$0.03 & 2084$\pm$156 & 7.55$\pm$0.07 & $-2.04^{+0.70}_{-0.93}$ & $2.00^{+1.12}_{-0.82}$ \\
			1088832 & 189.3443 & 62.2634 & 5.24 & WFSS & 43.10$\pm$0.02 & 2256$\pm$91 & 7.85$\pm$0.04 & -1.27$\pm$0.20 & -- \\
			1013188 & 189.0571 & 62.2689 & 5.25 & WFSS & 42.33$\pm$0.03 & 1957$\pm$147 & 7.36$\pm$0.07 & -1.44$\pm$0.29 & -- \\
			1020514 & 189.1793 & 62.2925 & 5.36 & WFSS & 42.40$\pm$0.03 & 1612$\pm$140 & 7.22$\pm$0.08 & -1.91$\pm$0.07 & -0.32$\pm$0.37 \\
			1087388 & 189.2810 & 62.2473 & 5.54 & WFSS & 43.52$\pm$0.01 & 2965$\pm$55 & 8.29$\pm$0.02 & -0.30$\pm$0.19 & -- \\
			1011836 & 189.2206 & 62.2637 & 4.41 & NIRSpec & $41.86^{+0.03}_{-0.03}$ & $1451^{+98}_{-105}$ & 7.13$\pm$0.31 & -0.94$\pm$0.11 & -2.14$\pm$0.08 \\
			1020621 & 189.1225 & 62.2929 & 4.68 & NIRSpec & $42.00^{+0.03}_{-0.03}$ & $1638^{+148}_{-150}$ & 7.30$\pm$0.31 & -1.18$\pm$0.17 & 0.88$\pm$0.33 \\
			1001093 & 189.1797 & 62.2246 & 5.60 & NIRSpec & $41.86^{+0.04}_{-0.03}$ & $1662^{+203}_{-165}$ & $7.36^{+0.32}_{-0.31}$ & -1.30$\pm$0.30 & 0.12$\pm$0.23 \\
			1061888 & 189.1680 & 62.2170 & 5.87 & NIRSpec & $42.15^{+0.03}_{-0.02}$ & $1375^{+97}_{-127}$ & 7.22$\pm$0.31 & -1.96$\pm$0.17 & -- \\
        \hline
		\end{tabular}
        \caption{The broad-line selected AGN sample in this work. \textsc{ID} refers to the source ID in the JADES DR3 photometric catalog \citep{JADES_DR3}, and $z_{\rm spec}$ is the spectroscopic redshift measured from the NIRCam WFSS data. (ID 9994014 is not included in the JADES DR3 catalog because of a diffraction-spike mask, but is reported in \citealt{Matthee2024}.) \textsc{Selection} indicates the different selection criteria outlined in \S\ref{sec:agn_sample}. The broad \ha\ luminosities ($L_{\rm H\alpha, broad}$), full widths at half maximum of the broad \ha\ emission lines (FWHM$_{\rm H\alpha, broad}$), and black hole masses ($M_{\rm BH}$) for the WFSS-selected AGNs are taken from \citet{Junyu2024}, while those for the NIRSpec-selected AGNs are from \cite{Maiolino2023}. The UV ($\beta_{\rm UV}$) and optical ($\beta_{\rm opt}$) slopes are calculated using \textsc{kron\_s} photometry (Kron radius = 1.2) with at least two strong line-free bands spanning $>1000\ \text{\AA}$ in the rest frame. 
    \label{tab:broadline_agn}}
\end{table*}

\begin{table*}[!ht]
	\begin{center}
		\begin{tabular}{ccccccc}
        \hline
			ID & RA & DEC & $z_{\rm spec}$ & selection & $\beta_{\rm UV}$ & $\beta_{\rm opt}$ \\
			\hline
			1055902 & 189.2348 & 62.2000 & 4.02 & V-shaped SED & -1.82$\pm$0.19 & 0.05$\pm$0.30 \\
			1029892 & 189.0928 & 62.2662 & 4.47 & V-shaped SED & -0.85$\pm$0.23 & 0.14$\pm$0.29 \\
			1066100 & 189.2272 & 62.2341 & 4.76 & V-shaped SED & -1.60$\pm$0.27 & 0.79$\pm$0.31 \\
			1081040 & 189.2816 & 62.2161 & 4.76 & V-shaped SED & -1.71$\pm$0.20 & 1.30$\pm$0.25 \\
			1020485 & 189.1131 & 62.2924 & 5.27 & V-shaped SED & -1.70$\pm$0.23 & 0.60$\pm$0.27 \\
        \hline
		\end{tabular}
        \caption{The V-shaped SED-selected AGNs in this work, as a complement to the broad-line sample in Table \ref{tab:broadline_agn}. \textsc{ID} refers to the source ID in the JADES DR3 photometric catalog, and $z_{\rm spec}$ is the spectroscopic redshift measured from the NIRCam WFSS data. 
        \label{tab:vshaped_agn} }
	\end{center}
\end{table*}

\bigskip

\section{Diverse environments of JWST-selected high-redshift AGNs}\label{sec:DiverseEnv}

In this section, we study the large-scale environments of  AGNs at $z=3.9-6$  based on the grism-selected HAEs.  

\subsection{Overdensity field $\delta$ of high-redshift AGNs}\label{sec:delta}

We calculate the overdensity field, $\delta$, for each AGN based on the number density of HAEs within a volume of (15 $h^{-1}$ cMpc)$^3$, which equals the volumes defined in \cite{Chiang2013} and has been widely adopted in the literature \citep[e.g.,][]{Chiang2014, Cucciati2014, Cai2016, Helton2024, Lim2024}. We center a cylinder with a radius of 8.5 $h^{-1}$ cMpc and a length of 15 $h^{-1}$ cMpc on the AGN, and calculate the galaxy density, $n$, within its volume by counting all enclosed galaxies. The WFSS footprint edge is taken into account. To calculate the mean galaxy density, $\bar{n}$, we use the random galaxy catalog from the HAE autocorrelation analysis in \cite{Lin_GDN_HAE}. The random galaxy catalog is designed to simulate a uniform distribution of galaxies without any clustering signal, while following the same selection function as the observed sample. The line fluxes of random galaxies are drawn from the \ha\ luminosity functions, and the completeness model is applied to account for selection effects.  We count the random galaxies within the same volume and then scale the value to match the total HAE number density within the field of view (FoV).  The overdensity field is computed as $\delta=n/\bar{n}-1$.  The uncertainty of $\delta$ includes Poisson fluctuations of the observed galaxies and dispersion in the random galaxy distribution across different random galaxy realizations.

\begin{figure*}[htbp!]
    \centering    \includegraphics[width=\linewidth]{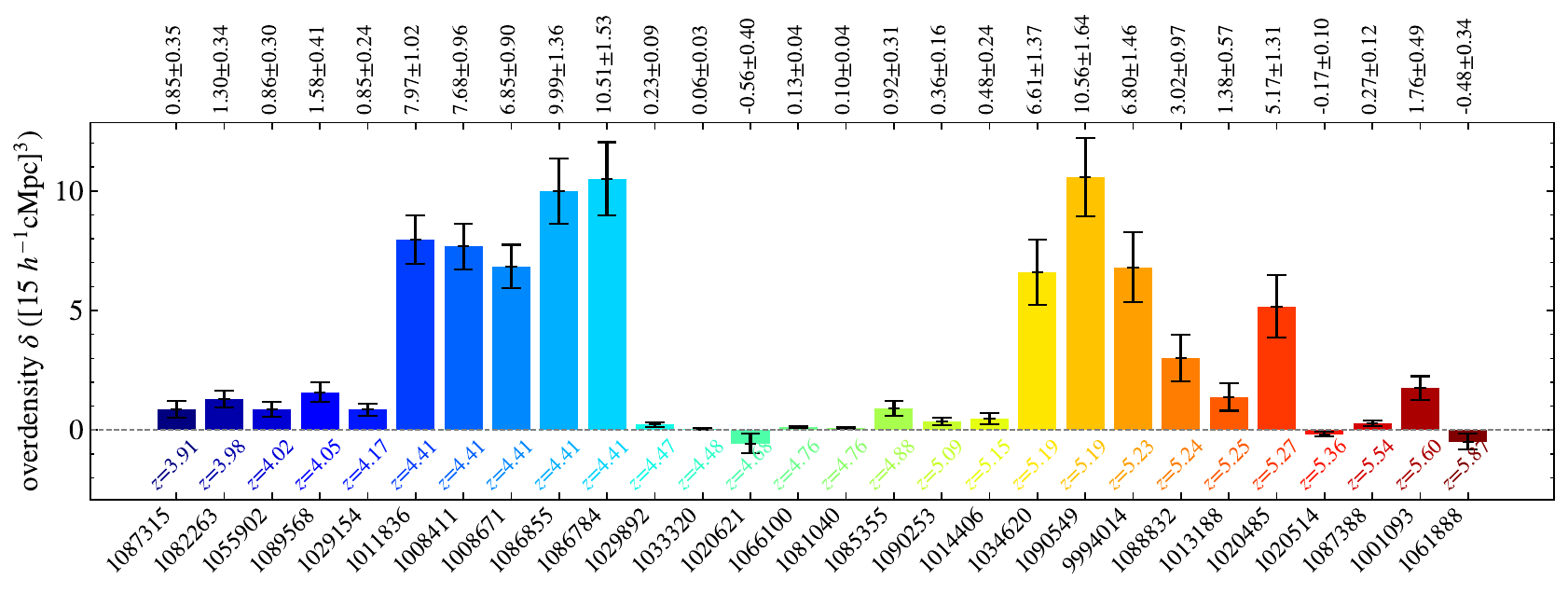}
    \caption{The overdensity fields $\delta$ over a volume of (15 $h^{-1}$ cMpc)$^{3}$ for AGNs at $3.9<z<6$.  The AGNs are ordered and color-coded by their redshift, with their IDs shown on the $x$-axis. The $\delta$ value is labeled at the top of each AGN.}
    \label{fig:AGN_delta}
\end{figure*}

\begin{figure*}[htbp!]
    \centering
\includegraphics[width=0.454\textwidth]{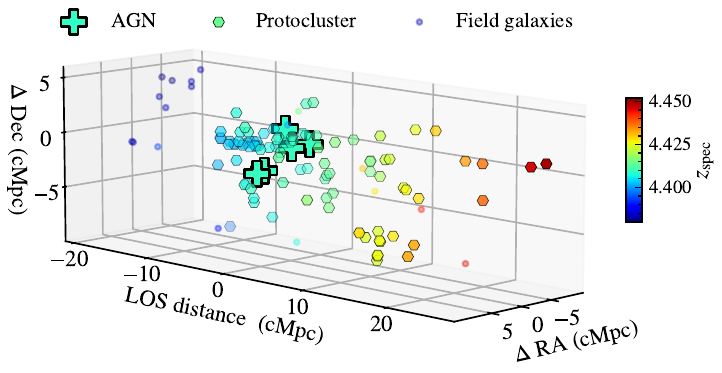}
\includegraphics[width=0.54\textwidth]{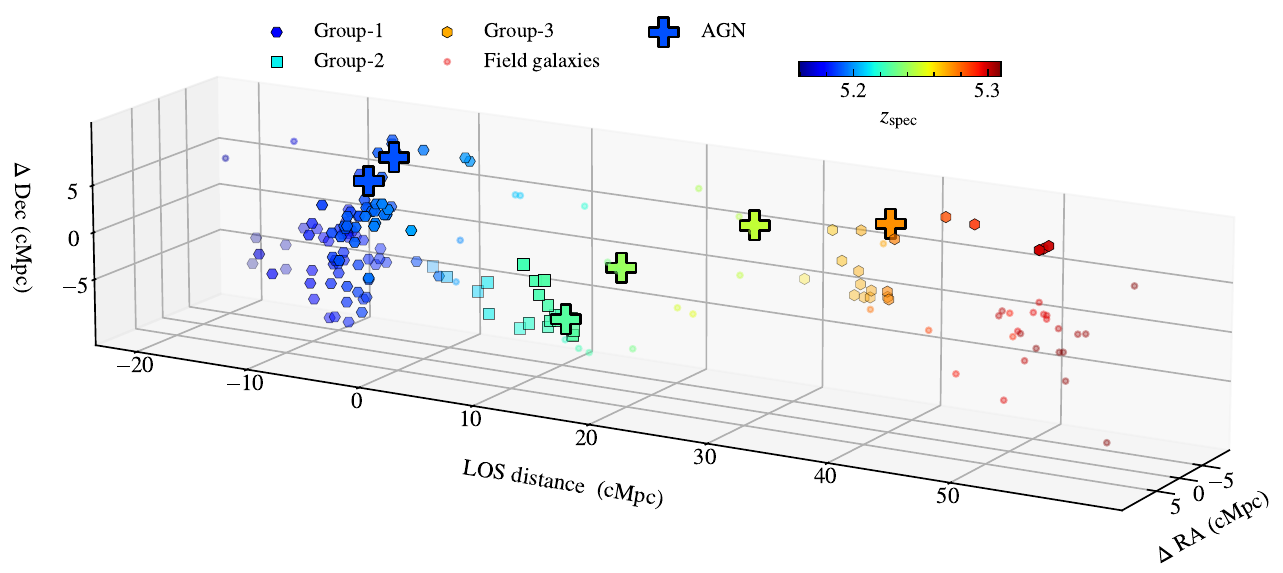}
    \caption{The 3D large-scale structure of the protocluster at $z\approx4.41$ (left) and $z\approx5.19$ (right), after the luminosity cut of $10^{41.5}$ \si{erg\,s^{-1}} and grouping within 500 pkpc and 500 \si{km\,s^{-1}}. They host five and six AGNs respectively. The coordinates of the $z\approx4.41$ structure are with respect to (RA, DEC, $z$)=(189.21355, 62.24861, 4.41), and the coordinates of the $z\approx5.19$ structure are
    with respect to (RA, DEC, $z$)=(189.20403, 62.23787, 5.195). All sources are color-coded by their redshifts, with the plus indicating AGNs, hexagons/squares indicating the protocluster members, and dots indicating field galaxies. }
    \label{fig:protocluster_structure}
\end{figure*}

The overdensity $\delta$ around AGNs is shown in Figure \ref{fig:AGN_delta}.  The JWST AGNs are located in highly diverse large-scale environments.  Their $\delta$ values range from $-0.56$, indicating underdense environments, to $10.56$, representing an extreme overdensity. Among them, 10 AGNs ($36 \pm 9$\% assuming a binomial distribution) are in overdense regions with $\delta>3$, while the remaining 18 AGNs ($64 \pm 9$\%) are in average or underdense regions. We identify two extreme protoclusters at $z = 4.41$ and $z = 5.19$. The protocluster at $z \approx 4.41$ contains five AGNs and 92 HAEs after applying the luminosity cut and grouping. These AGNs show overdensities ranging from $\delta = 6.85$ to $10.51$.  Six AGNs at $5.16 < z < 5.29$ are found in the filamentary structure at $z = 5.19$ \citep{Herard_Demanche_2023, Sun2024}. The structure consists of three galaxy groups, with 93 galaxies at  $z=5.19$,  22 galaxies at $z=5.22$, and 19 galaxies at $z=5.27$. Among the six AGNs, two are associated with the $z = 5.19$ galaxy group, and two are associated with the $z = 5.22$ and $z = 5.27$ galaxy groups. These four exhibit $\delta = 5.17 - 10.56$. The remaining two AGNs reside between these groups, with $\delta = 3.02$ and $1.38$, respectively. We show the 3D structures of both protoclusters in Figure \ref{fig:protocluster_structure}.

As a baseline for comparison, we also calculate the $\delta$ values for HAEs at similar redshifts. Among the \NHAE\ HAEs at $3.9 < z < 6$, 232 ($30\pm 2\%$) lie in regions with $\delta > 3$. The fraction of galaxies in overdense regions is broadly consistent with that of the AGNs within $1\sigma$. We conclude that, statistically, AGNs in our sample are not preferentially located in denser large-scale environments compared to star-forming galaxies.

\subsection{Dependence of AGN properties on their environments}

\begin{figure*}[htbp!]
    \centering
    \includegraphics[width=\linewidth]{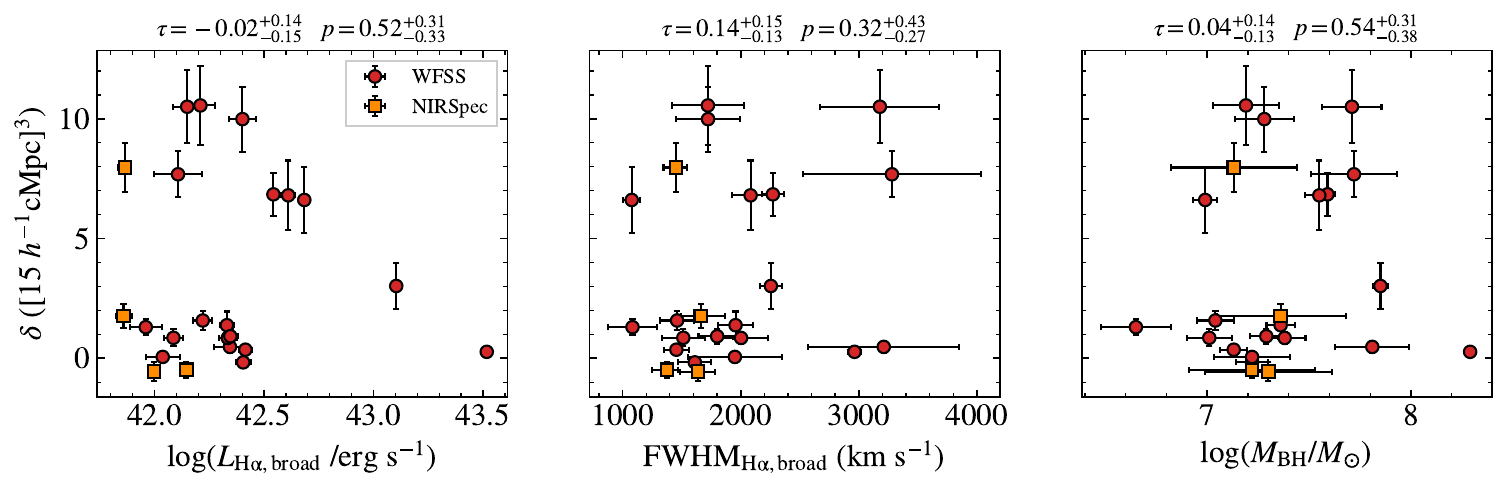}
    \caption{The overdensity fields $\delta$ of AGNs vs. their broad \ha\ luminosities (left), FWHMs of their broad \ha\ lines (middle), and BH masses (right). The red dots represent the WFSS-selected broad-line AGNs and the orange squares are the NIRSpec-selected broad-line AGNs. The Kendall's $\tau$ analysis, with $\tau$ and $p$ labeled on top of each panel, suggests no significant correlation between $\delta$ and the three parameters. }
    \label{fig:delta_AGN_property}
\end{figure*}

We explore the correlation between $\delta$ and the broad-line \ha\ luminosity ($L_{\rm H\alpha, broad}$), the FWHMs of the broad \ha\ (FWHM$_{\rm H\alpha, broad}$), and the BH mass (\MBH). We use only the broad-line selected AGNs from the grism and NIRSpec \citep{Junyu2024, Maiolino2023}, for which these parameters can be measured. We note that precise \MBH\ values are uncertain due to the unclear nature of these low-luminosity AGNs. We follow the common practice of deriving BH masses from $L_{\rm H\alpha, broad}$ and FWHM$_{\rm H\alpha, broad}$ to examine the potential correlation. In this work, \MBH\ is calculated following \cite{Reines2015}, as a result of $L_{\rm H\alpha, broad}$ and FWHM$_{\rm H\alpha, broad}$ ($M_{\rm BH}\propto {\rm FHWM^{2.06}_{H\alpha, broad}}L^{0.47}_{\rm H\alpha, broad}$).  We perturb these parameters within their uncertainties to estimate the uncertainties in the correlation coefficient\footnote{We use \href{https://github.com/privong/pymccorrelation}{\texttt{pymccorrelation}} to implement the perturbation \citep{Curran2014, Privon2020}.}. The correlations between $L_{\rm H\alpha, broad}$, FWHM$_{\rm H\alpha, broad}$, and \MBH\ with $\delta$ are shown in Figure \ref{fig:delta_AGN_property}. No clear correlation is found for $L_{\rm H\alpha, broad}$, FWHM$_{\rm H\alpha, broad}$, or \MBH, as the Kendall's $\tau$ analysis yields $p > 0.05$.\footnote{The $p$ value denotes the probability of obtaining the current result if the correlation coefficient were zero (no correlation). If $p$ is lower than 0.05, the correlation coefficient is statistically significant.}

\begin{figure}[htbp]
    \centering
    \includegraphics[width=\linewidth]{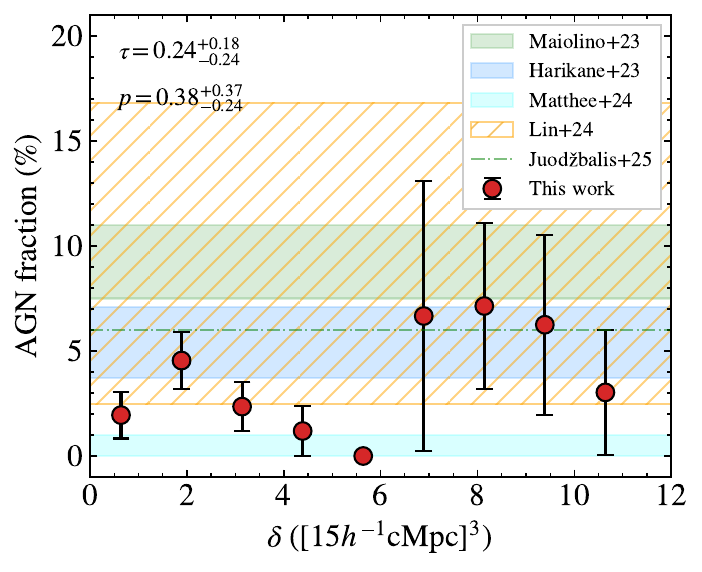}
    \caption{AGN fraction among the star-forming galaxies vs. the overdensity field $\delta$. The Kendall's $\tau$ statistic and the associated probability $p$ are labeled at the top left, revealing no significant correlation between AGN fraction and $\delta$. The AGN fractions from literature studies are shown as green, blue, cyan, and orange shaded regions and green dashed-dotted lines for reference \citep{Maiolino2023, Harikane2023, Matthee2024, Lin2024, Juodvzbalis2025}. Note that these different AGN fractions are luminosity-dependent and subject to cosmic variance and selection effects. }
    \label{fig:AGN_fraction}
\end{figure}

We define the AGN fraction in this work as the proportion of AGNs that meet any of the three selection criteria (\S\ref{sec:agn_sample}) among all grism-selected HAEs. The averages are 3.7\% at $3.9 < z < 5$,  4.0\%  at $5 < z < 6$, and  3.9\%  for the combined range $3.9 < z < 6$.  We calculate the overdensity, $\delta$, for all HAEs in the range $3.9 < z < 6$ using the same method described in \S\ref{sec:delta}, and estimate the AGN fraction within each fixed $\delta$ range. Figure \ref{fig:AGN_fraction} shows the AGN fraction as a function of $\delta$. The uncertainties are calculated assuming a binomial distribution for the AGN fraction.  The Kendall's $\tau$ analysis implies no significant dependence of AGN fraction on $\delta$, with a correlation coefficient $\tau$ consistent with 0 and a $p$-value much greater than 0.05. We note that the AGN fractions shown here do not include corrections for selection functions or completeness, particularly for those selected through NIRSpec. For comparison, in Figure \ref{fig:AGN_fraction} we also show AGN fractions compiled from NIRSpec broad-line selections \citep{Maiolino2023, Harikane2023, Juodvzbalis2025} and from grism-based broad-line selections \citep{Matthee2024, Lin2024}. Although the AGN fractions are luminosity-dependent and may be affected by cosmic variance, the absence of correlations in our results is less susceptible to these factors. A more accurate determination of the AGN fraction will require future observations with higher sensitivity, more uniform selection criteria, and careful accounting for selection effects.

The overdensity, $\delta$, on 15 $h^{-1}$cMpc scales is primarily driven by the large-scale structure dominated by the linear two-halo terms
\cite[e.g.,][]{Harikane2016, HerreroAlonso2023}. The lack of dependence of $\delta$ on $M_{\rm BH}$ implies significant scatter in the relation between large-scale environments and black hole mass in high-redshift AGNs.  
On the other hand, the bolometric luminosity and Eddington ratios can be derived using $L_{\rm H\alpha, broad}$ and FWHM$_{\rm H\alpha, broad}$ by applying the empirical relations for typical type-1 AGNs to these low-luminosity AGNs \citep{Greene2005, Trakhtenbrot2011}, although this may not be accurate but is commonly adopted \citep[e.g.,][]{Maiolino2023, Matthee2024, Lin2024}. In this context, the lack of correlation of $\delta$ with respect to $L_{\rm H\alpha, broad}$ and FWHM$_{\rm H\alpha, broad}$ suggests that bolometric luminosity and Eddington ratios do not correlate with $\delta$ either. These results suggest that the large-scale environment on scales $>$10 cMpc does not significantly affect black hole growth and AGN evolution on smaller, pc scales.

In contrast, \cite{Matthee2024b} finds that overdensity within 1 cMpc is positively correlated with $M_{\rm BH}$, particularly when including measurements from 10 high-redshift AGNs and five quasars \citep{Eilers2024}.  \cite{Lin2024} also reports tentative evidence of small-scale clustering around one low-luminosity AGN, with three neighboring galaxies located within 30 kpc. Our WFSS observations are not as deep as those in \cite{Matthee2024b}, which were conducted in a lensing cluster field with longer exposures, limiting our ability to probe overdensities within 1 cMpc. The 1 cMpc scale is around the typical transition point between the dominance by non-linear one-halo terms and linear two-halo terms for high-redshift galaxies \cite[e.g.,][]{Harikane2016, HerreroAlonso2023}. The differing environmental effects on $M_{\rm BH}$ at scales $<$1 cMpc and $>10$ cMpc suggest that black hole growth may be more strongly influenced by local non-linear processes such as mergers. Alternatively, processes occurring on cosmological scales, such as the gas accretion from the large-scale cosmic web, have a weaker impact. These large-scale activities may require a longer time to come into effect and impact the small-scale BH activities.  Future deep spectroscopic surveys are required to provide further insights by probing multi-scale environments of a large AGN sample across a wide range of $M_{\rm BH}$.

\bigskip

\section{The clustering analysis of high-redshift AGNs}\label{sec:clustering}

As illustrated in \S\ref{sec:DiverseEnv}, there is significant variance in the environments of high-redshift AGNs. We therefore investigate their average population-level clustering properties in the context of large-scale structure of star-forming galaxies at the same cosmic epochs. For this clustering analysis, we focus exclusively on the 26 AGNs that meet the grism broad-line selection and V-shaped SED criteria.  Two of the four NIRSpec-selected AGNs meet the V-shaped SED criteria and are therefore included in the clustering analysis. We do not include the remaining two NIRSpec-selected AGNs because of the limitation in quantifying their selection function via the Micro Shutter Array (MSA) design. The MSA targets must be pre-selected, resulting in a biased search.

\subsection{The projected surface density excess}\label{sec:surfacedensity}

We first compare the galaxy surface number density distributions centered on AGNs and HAEs. We define the
projected surface density excess within a radius of $r_p$ by

\begin{equation}
    \Sigma(<r_p) = \frac{N(<r_p)}{A(<r_p)} - \frac{N_{\rm all}}{A_{\rm all}},
\end{equation}
where $r_p$ is the projected distance on the sky plane. $N(<r_p)$ is the number of HAEs within a cylindrical volume of radius $r_p$ and line-of-sight length $8\, h^{-1} \, \mathrm{Mpc}$.  The line-of-sight length corresponds to approximately 1000 \si{km.s^{-1}} at $z \approx 5$. We have tested and confirmed that this length effectively captures the large-scale structure signal along the line-of-sight while preserving optimal signal-to-noise ratios for the clustering measurements. $A(<r_p)$ is the area on the projected sky plane within radius $r_p$. $N_{\rm all}$ and $A_{\rm all}$ are the total galaxy number and projected area of the WFSS footprint, respectively. The value $N_{\rm all}/A_{\rm all}$ is used to evaluate the average projected surface density across the entire survey area.
$\Sigma(<r_p)$ represents the excess in the galaxy number surface density within a radius of $r_p$, relative to the average surface density. 

\begin{figure}
    \centering
    \includegraphics[width=\linewidth]{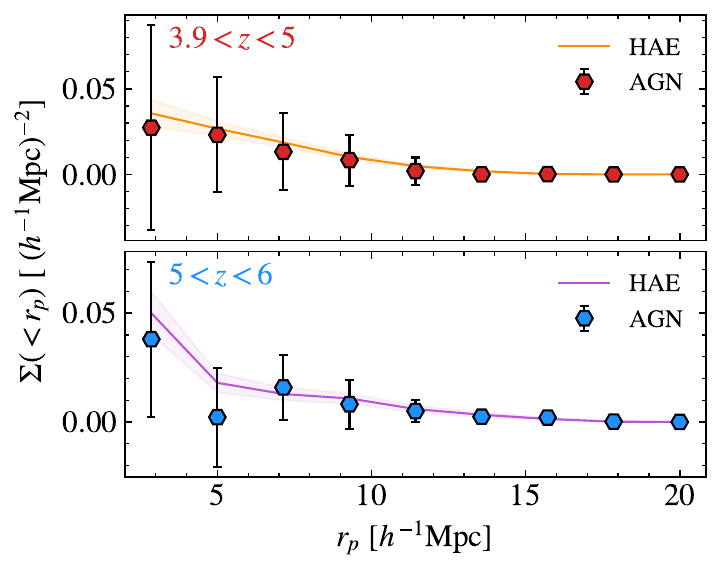}
    \caption{The projected surface density excess of AGNs compared to that of HAEs. The lines represent the median of $\Sigma(<r_p)$ for HAEs, with the shaded regions indicating the 1$\sigma$ range. The hexagons denote the median of $\Sigma(<r_p)$ for AGNs. The error bars indicate the 1$\sigma$ level of the $\Sigma(<r_p)$ values.}
    \label{fig:agn_sd}
\end{figure}

We calculate $\Sigma(<r_p)$ for each AGN and HAE. The median projected surface density excess of AGNs and HAEs at $3.9<z<5$ and $5<z<6$ is presented in Figure \ref{fig:agn_sd}.  The projected surface density excess of AGNs is consistent with that of HAEs, although it shows a large variance at $r_p < 10~h^{-1}$ Mpc. The variances arise from the diversity of AGN environments, as well as the limited survey volume and depth, which primarily dominate the uncertainty in the two bins at $r_p < 5~h^{-1}$ Mpc.
Nevertheless, the broad agreement in the projected surface density excess of AGNs and HAEs suggests that star-forming galaxies do not have stronger clustering around AGNs, compared to the normal galaxy-galaxy clustering.

\begin{figure*}[!ht]
    \centering
    \includegraphics[width=\linewidth]{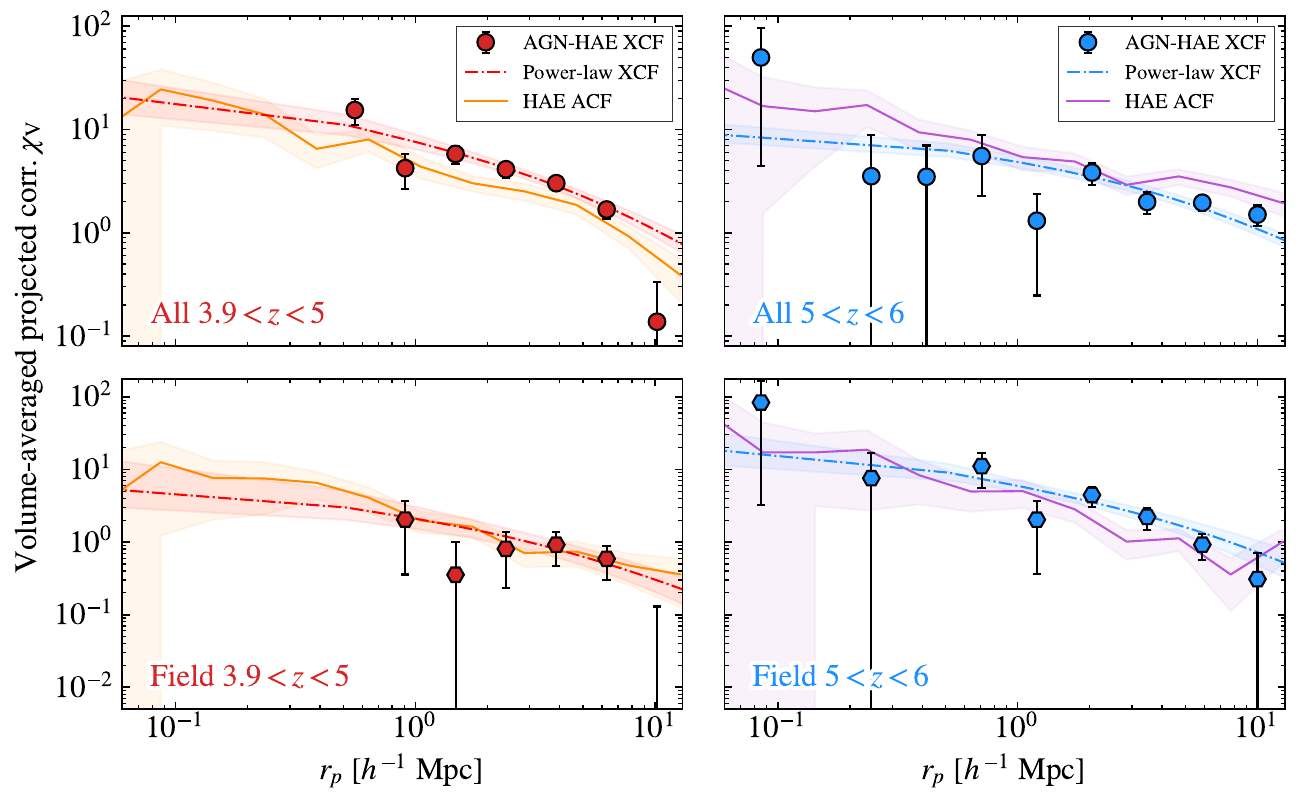}
    \caption{The volume-averaged AGN-HAE cross-correlation compared to the HAE autocorrelation. \textit{Top panels}: The correlation functions consider all HAEs and AGNs, including those in the protoclusters. In the \textit{top-left} panel, the red dots show the cross-correlation between AGNs and HAEs at $3.9<z<5$. The dashed-dotted red line and red shaded region denote the best-fit power-law model and its uncertainty. The orange line and orange shaded regions are the HAE autocorrelation and its uncertainty at $4\lesssim z<5$. In the \textit{top-right} panel, the blue dots show the cross-correlation between AGNs and HAEs at $5< z<6$, with the dashed-dotted blue line indicative of the best-fit power-law model. The purple line and purple shaded regions are the HAE autocorrelation and its uncertainty at $5< z<6$. \textit{Bottom panels}: Similar to the \textit{top} panels, but for AGNs and HAEs in the field. We exclude the $z\approx 4.41$ and $z\approx 5.19$ protoclusters from both the AGN-HAE cross-correlation and HAE autocorrelation. }
    \label{fig:AGN_HAE_xcf}
\end{figure*}

\begin{table*}[!ht]
\begin{center}
\begin{tabular}{cccccc}
\hline\hline
 & & \multicolumn{2}{c}{\textsc{all}} & \multicolumn{2}{c}{\textsc{field}} \\
Parameter &  & $3.9<z<5$ & $5<z<6$ & $3.9<z<5$ & $5<z<6$ \\
\hline
$\gamma$  & & $1.27_{-0.15}^{+0.16}$ & $1.08_{-0.06}^{+0.10}$ & $1.35_{-0.25}^{+0.42}$ & $1.30_{-0.20}^{+0.27}$ \\
$r_0$ & ($h^{-1}$Mpc) & $12.37_{-1.70}^{+2.30}$ & $11.83_{-1.37}^{+1.58}$ & $4.26_{-0.70}^{+0.83}$ & $9.94_{-2.25}^{+3.06}$ \\
$\gamma$ (fixed)  & & $1.34$ & $1.02$ & $1.59$ & $1.63$ \\
$r_0$ ( $\gamma$ fixed) & ($h^{-1}$Mpc) & $11.75_{-0.87}^{+0.94}$ & $12.63_{-1.35}^{+1.45}$ & $4.26_{-0.61}^{+0.70}$ & $7.66_{-0.81}^{+0.90}$ \\
$r^{gg}_0$ & ($h^{-1}$Mpc) & $8.29_{-1.21}^{+1.66}$ & $19.71_{-2.90}^{+3.38}$ & $4.61_{-0.68}^{+1.00}$ & $6.23_{-1.13}^{+1.68}$ \\
$\gamma^{gg}$ & & $1.34_{-0.14}^{+0.13}$ & $1.02_{-0.02}^{+0.03}$ & $1.59_{-0.25}^{+0.23}$ & $1.63_{-0.26}^{+0.24}$ \\
$\log M_{h}$ & ($M_\odot$) & $11.21_{-0.33}^{+0.36}$ & $11.06_{-0.32}^{+0.36}$ & $11.21_{-0.32}^{+0.35}$ & $11.04_{-0.32}^{+0.34}$ \\
$\log M_{\rm *}$ & ($M_\odot$) & $8.57_{-0.71}^{+0.80}$ & $8.41_{-0.70}^{+0.79}$ & $8.55_{-0.73}^{+0.77}$ & $8.36_{-0.70}^{+0.74}$ \\
$b_g$ & & -- & -- & $4.11\pm0.12$ & $5.90\pm0.08$ \\
$b_a$ & & -- & -- & $2.95\pm0.99$ & $8.19\pm1.81$ \\
$\log M_{h, b_g}$ & ($M_\odot$) & -- & -- & $11.20\pm0.05$ & $11.25\pm0.02$ \\
$\log M_{h, b_a}$ & ($M_\odot$) & -- & -- & $10.57_{-0.97}^{+0.56}$ & $11.76^{+0.26}_{-0.38}$ \\
\hline
\end{tabular}
\end{center}
	\caption{The best-fit power-law parameters for the AGN-HAE cross-correlations presented for two cases: including AGNs and galaxies within the protoclusters (\textsc{all}) and including only the field AGNs and galaxies (\textsc{field}). The parameters $r_0$ and $\gamma$ represent the characteristic scale length and the slope of the AGN-HAE cross-correlation, respectively. For comparison, we also provide the scale length  ($r_0^{gg}$) and slope  ($\gamma^{gg}$) of the HAE autocorrelations, along with the derived halo masses ($M_h$) and stellar masses ($M_\star$) from \textsc{UniverseMachine} \citep{Behroozi2019}. The values represent the median of the $M_h$ and $M_\star$ distributions. The $1\sigma$ ranges are based on the 16th and 84th percentiles of these distributions. \xj{As a complement, we also present the measured bias for field HAEs ($b_g$) and AGNs ($b_a$), and the corresponding halo masses derived from the biases ($M_{h,b_g}$,  $M_{h,b_a}$).}  }
	\label{tab:agn_hae_xcf}
\end{table*}

\subsection{The cross-correlation between \ha\ emitters and AGNs}\label{sec:agn_xcf}

Because the sample size of AGN is too small for autocorrelation analysis, we instead study the cross-correlation between HAEs and AGNs to examine the clustering of AGNs.
We adopt the LS estimators \citep{LS} as follows: 
 
\begin{equation}
\begin{split}
\xi_{ag}(\vec{r}) = &\frac{N_{\mathrm{r}}\left(N_{\mathrm{r}}-1\right)}{2 N_{\mathrm{a}} N_{\mathrm{g}} } \frac{AG (\vec{r})}{R R(\vec{r})} 
- \frac{N_{\mathrm{r}}-1}{2 N_{\mathrm{g}}} \frac{G R(\vec{r})}{R R(\vec{r})} \\
&- \frac{N_{\mathrm{r}}-1}{2 N_{\mathrm{a}}} \frac{A R(\vec{r})}{R R(\vec{r})} + 1,
\end{split}
\end{equation}
where $N_a$ is the number of AGNs, $AG$ represents the number of AGN-HAE pairs and $AR$ represents the number of AGN-random pairs.  We compute the volume-averaged projected $\chi_V$ following \cite{Hennawi2006}: 
\begin{equation}\label{eq:chiV}
\chi_V\left(r_p\right)=\frac{2}{V} \int_{r_{p, \min }}^{r_{p, \max }} \int_0^{r_{\pi,\max }} \xi\left(r_p, r_\pi\right) 2 \pi r_p \mathrm{~d} r_p \mathrm{~d} r_\pi,
\end{equation}
where $r_{\pi, \max}$ is the integration limit along the line-of-sight direction. We set $r_{\pi, \max}$ to be 8 $h^{-1}$cMpc, corresponding to approximately 1000 \si{km\,s^{-1}} at $z\approx 4-6$ and can effectively capture the clustering signals of large-scale structure. $V$ is the volume of the cylindrical shell in the $r_p$ bin ($r_{p, \min }<r_p<r_{p, \max }$), which can be expressed as  $V=\pi \left(r_{p, \max }^2-r_{p, \min }^2\right) \cdot r_{\pi, \max}$. We construct the covariance matrix using bootstrapping. Each time, we sample the AGN and HAE catalogs with replacement and calculate the cross-correlation using different random catalog realizations. The uncertainties are derived from the diagonal of the covariance matrix.

The measured cross-correlations for all AGNs and HAEs at $3.9 < z < 5$ and $5 < z < 6$, including those in protoclusters, are shown in Figure \ref{fig:AGN_HAE_xcf}. We also calculate $\chi_V$ for field AGNs and HAEs by excluding the two protoclusters at $z=4.41$ and 5.19 shown in Figure \ref{fig:protocluster_structure}. We
parameterize the correlation functions by a power law:
\begin{equation}
\xi(r)=\left(r / r_0\right)^{-\gamma},
\end{equation}
where $r_0$ denotes the characteristic scale length (i.e., $\xi(r_0)=1$). The best-fit power-law models are listed in Table \ref{tab:agn_hae_xcf}. For comparison, we also list the best-fit parameters for the field HAE autocorrelation functions and those of all HAEs including protocluster member galaxies. The HAE clustering analysis is described in \citet{Lin_GDN_HAE}.  While the power-law slope $\gamma$ for field AGN-HAE cross-correlations is consistent with the typical range of $1.6 - 2.0$ reported in the literature \citep[e.g.,][]{Hennawi2006, Geach2012, Eilers2024}, the slopes of AGN-HAE cross-correlations including protocluster members are flatter ($\gamma = 1.27, 1.06$ at $z\approx3.9-5$ and $z\approx5-6$, respectively). These flat $\gamma$ values are also observed in the HAE autocorrelations with protocluster galaxies. The flattening is attributed to the filamentary geometry of protoclusters, which is clearly reflected in the 2D $\xi(r_p, r_\pi)$ planes. We refer interested readers to \citet{Lin_GDN_HAE} for a more detailed discussion of the geometry effect. To better compare the amplitudes of AGN-HAE cross-correlations with those of HAE autocorrelations, we fix the $\gamma$ of the AGN-HAE cross-correlations to match that of the HAE autocorrelations.

The AGN-HAE cross-correlation at $3.9 < z < 5$, when including sources from the $z \approx 4.41$ protocluster, exhibits a higher amplitude than the HAE autocorrelation. This is due to the higher fraction of AGNs (5 out of 16, or 31\%) residing in the protocluster compared to HAEs (92 out of 482, or 19\%). Despite this, the two amplitudes remain consistent within $2\sigma$.  At $5 < z < 6$, the AGN-HAE cross-correlation, when including the $z = 5.19$ protocluster, has a lower amplitude than the HAE autocorrelation. Overall, variations in the AGN-HAE cross-correlation that include all HAEs are primarily driven by cosmic variance introduced by the two protoclusters.  In contrast, at $3.9 < z < 5$ and $5 < z < 6$ (bottom panel of Figure \ref{fig:AGN_HAE_xcf}), the field AGN-HAE cross-correlation amplitudes are consistent with those of the field HAE autocorrelation. When both correlation functions are fitted with the same power-law slope ($\gamma$), their amplitudes agree within $1\sigma$ (Table \ref{tab:agn_hae_xcf}).  We conclude that the AGN-HAE cross-correlation has an amplitude comparable to that of the HAE autocorrelation. This consistency in the correlation function aligns with the similar amplitude of the projected surface density excess for AGNs and HAEs, as discussed in \S\ref{sec:surfacedensity}.

Based on the linear perturbation theory (see an overview about galaxy bias in \citealt{Desjacques2018}), the two-point correlation functions for HAEs, \(\xi_{gg}(\vec{r})\), and for AGNs and HAEs, \(\xi_{ag}(\vec{r})\), can be expressed as
\begin{align}\label{eq:xi_aagg}
    \xi_{ag}(\vec{r})^2 &= \xi_{aa}(\vec{r})\xi_{gg}(\vec{r}) \\
    &= b_a b_g \xi_{mm}^2(\vec{r}),   \label{eq:xi_ag_bias} 
\end{align}
where $b_g$ and $b_a$ are the bias factors for the enhancement of the perturbation in HAE and AGN number densities, and $\xi_{mm}(\vec{r})$ is the dark matter autocorrelation function.  According to Equation \ref{eq:xi_aagg}, the comparable amplitudes of $\xi_{ag}$ and $\xi_{gg}$ suggest comparable amplitudes of $\xi_{aa}$ and $\xi_{gg}$, and thus $b_g$ and $b_a$ should be similar. According to the linear bias theory, 
we conclude that AGNs and HAEs in our sample reside in dark matter halos of similar mass.

\bigskip

\begin{figure*}[!t]
    \centering
    \includegraphics[width=0.8\linewidth]{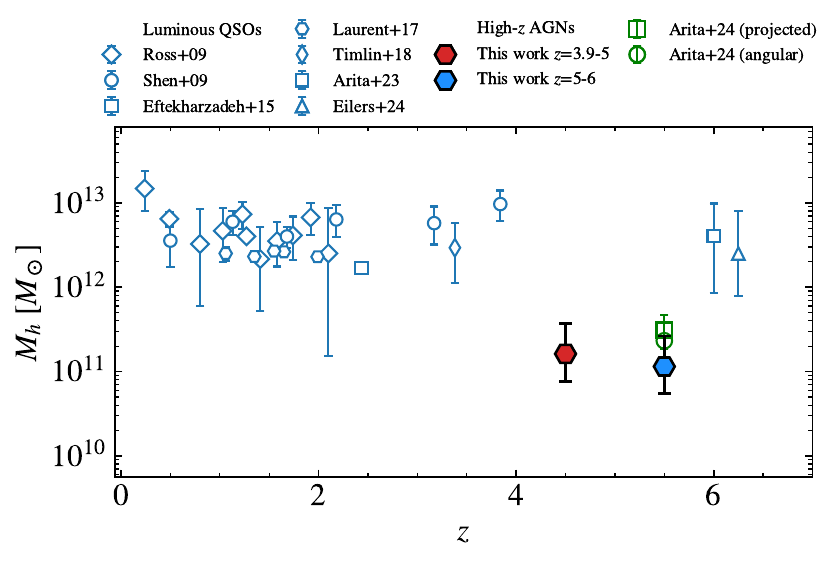}
    \vspace*{-4mm}
    \caption{The redshift evolution of dark matter halo masses for quasars and AGNs. For the bias measurements reported in the literature \citep{Ross2009, Shen2009, Eftekharzadeh2015, Laurent2017, Timlin2018, Arita2023, Arita2024}, we convert the bias values using our adopted cosmological parameters and the halo bias models from \cite{Tinker2010}. The $M_h$ value from \cite{Eilers2024} represents the median halo mass of quasars derived based on \textsc{universemachine}. We show two $M_h$ values from \cite{Arita2024} in green, derived based on the projected cross-correlation functions of low-luminosity AGNs and photometrically selected galaxies, and their angular cross-correlation, respectively.}
    \label{fig:hm_redshift}
\end{figure*}

\section{Discussion}\label{sec:um}

\subsection{Implications for the AGN host dark matter halo masses}
The similar amplitudes of the HAE autocorrelation function, $\xi_{gg}$, and the AGN-HAE cross-correlation function, $\xi_{ag}$, suggest that high-redshift AGNs in our sample reside in dark matter halos with masses comparable to those of star-forming galaxies at the same redshifts. In \citet{Lin_GDN_HAE}, we derive the dark matter halo masses ($M_h$) of HAEs by comparing the HAE autocorrelation with \textsc{UniverseMachine} simulations \citep{Behroozi2019}. The comparison accounts for both sample and cosmic variance, including the impact of large-scale structure geometry on the power-law shape of clustering.  We adopt the $M_h$ values of HAEs for AGNs, as shown in Table \ref{tab:agn_hae_xcf}. \xj{The $M_h$ values are taken as the medians of the simulated halo mass distributions, with the 16th and 84th percentiles indicating the scatter.} Note that we adopt $M_h$ defined for (sub)halos. Each (sub)halo hosts an individual galaxy, whether central or satellite, which cannot be distinguished based on our observations.  
The derived halo mass $M_h$ is approximately $10^{11} M_\odot$ for both protocluster members and field galaxies.  

\xj{As a complement, we estimate the bias for both HAEs and AGNs. Since the correlation functions of HAEs and AGNs in the full sample are affected by the protocluster geometry, we restrict the measurements to the field sample. The biases are derived by fitting the measured correlation functions at $r_p>1\,h^{-1}{\rm Mpc}$ to that of the underlying dark matter field, assuming the transfer function model from \textsc{camb} \citep{Lewis2011} and bias model of \cite{Tinker2010}. We first obtain the bias for HAEs ($b_g$) from the HAE autocorrelation function, fix it, and then fit the bias for AGNs ($b_a$) following Equation \ref{eq:xi_ag_bias}. The results are reported in Table~\ref{tab:agn_hae_xcf}. Despite the large uncertainties due to the limited sample size, the halo masses of field AGNs inferred from \textsc{UniverseMachine} ($M_h$) and from the bias measurements ($M_{h,b_a}$) are consistent within $2\sigma$. We adopt the \textsc{UniverseMachine} estimates as fiducial. We note that, although available only for the field sample, adopting the bias-derived values does not alter the conclusions below.}

We compare the $M_h$ of low-luminosity AGNs in our sample with luminous quasars across different cosmic epochs in Figure \ref{fig:hm_redshift}. For the clustering analysis of quasars or AGNs from large-area surveys (e.g., Sloan Digital Sky Survey, Subaru) with reported bias \citep[][]{Ross2009, Shen2009, Eftekharzadeh2015, Laurent2017, Timlin2018, Arita2023, Arita2024}, we convert their measured bias values into $M_h$ using our adopted cosmological parameters and the halo bias models of \cite{Tinker2010} with the \textsc{colossus}\footnote{\url{https://bdiemer.bitbucket.io/colossus/index.html}} package \citep{Diemer2018}. \xj{We also present the median halo mass for quasars in \cite{Eilers2024}, which is derived by searching for analog systems to the quasar
fields in \textsc{universemachine}.}  The bias in \cite{Arita2024} from the projected (angular) correlation analysis of 28 low-luminosity broad-line AGNs with photometrically selected galaxies corresponds to $\log(M_h / M_\odot) = 11.5 \pm 0.2 \, (11.4^{+0.2}_{-0.3})$, based on the conversion in this study. These values are consistent with our measurements, accounting for sample variance and systematic uncertainties.  In contrast, the host halo masses of luminous quasars at $z = 0$–6 typically exceed $10^{12} M_\odot$, 1–2 dex higher than those of low-luminosity AGNs, while with little dependence on the quasar luminosity \citep[e.g.,][]{Shen2009}. 

This implies that the differences between low-luminosity AGNs and UV-luminous quasars are not mainly caused by the obscuration from different geometries \citep{Netzer2015}. These newly discovered AGNs are not simple counterparts of UV-bright quasars in dust-enshrouded environments. 
The low-luminosity AGNs discovered by JWST may be inherently distinct from luminous quasars. 
Instead, they reside in considerably less massive dark matter halos. Given the comparable halo masses of AGNs and HAEs, along with the high AGN fraction of up to $\gtrsim$17\% at $L_{\rm H\alpha} \approx 10^{43}\ \rm{erg~s^{-1}}$ \citep{Lin2024}, low-luminosity AGNs might represent either a common stage in galaxy evolution or a distinct phase in the BH-galaxy co-evolution as discussed in \cite{Pizzati2024}. 

To explain the characteristics of low-luminosity AGNs, such as their weak X-ray/radio emission and variability, super-Eddington accretion mode has been proposed in many studies \citep[e.g.,][]{Greene2024, Inayoshi2024b, Lambrides2024, Mazzolari2024, Maiolino2025}. Super-Eddington accretion is also proposed, if these AGNs have low duty cycles  ($\lesssim 1\%$), to reconcile their BH masses \citep{Pizzati2024}. In this case, the predicted host halo masses would be $M_h \sim 10^{11} M_\odot$.  
Interestingly, this predicted $M_h$ value is consistent with our measurements and $M_h$ derived from an independent simulation suite. \xj{Since low-luminosity AGNs and star-forming galaxies occupy halos of comparable mass, the AGN fraction (3.6\% in our sample) provides a rough estimate of the duty cycle. This number is broadly consistent with \citet{Pizzati2024}.
}

Multiwavelength observations are essential for better characterizing this population and its accretion mode, including their stellar component, radiation field, and dust content. Large-volume surveys are required to probe the multiscale environments of BH growth across a broad range of BH masses.

\begin{figure}[!t]
    \centering
    \includegraphics[width=\linewidth]{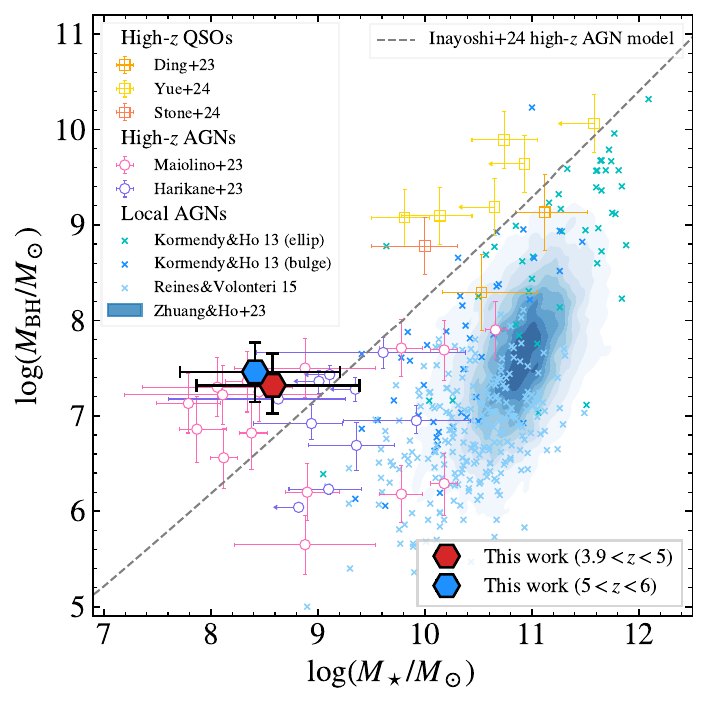}
    \caption{$M_{\rm BH}-M_\star$ relation. The orange and purple hexagons are positioned at the median $M_{\rm BH}$ measured from broad \ha\ lines, and median $M_\star$ for HAEs at similar redshift. The error bars represent the 16\%-84\% range. The literature $M_{\rm BH}$ and $M_\star$ values of quasars, high-redshift AGNs, and local AGNs are compiled from \cite{Ding2023, Yue2024, Stone2024, Maiolino2023, Harikane2023, Kormendy2013, Reines2015, Zhuang2023}. The model from \cite{Inayoshi2024} is shown as the dashed gray line.
    }
    \label{fig:Mbh_Ms_relation}
\end{figure}

 \subsection{Implication for the host galaxy stellar masses}

 Assuming the same stellar-to-halo mass ratio distribution, the stellar masses of AGN host galaxies should also be comparable to those of HAEs in the same redshift range.  The HAE stellar mass $M_\star$, derived using the  the \textsc{Univermachine} simulations, has a median value of $10^{8.6} M_\odot$ for $3.9 < z < 5$, and $10^{8.4} M_\odot$ for $5 < z < 6$, with $1\sigma$ scattering of approximately 0.8\,dex in both cases.   We note that the $M_\star$ distributions derived here are in good agreement with the $M_\star$ values obtained from the SED fitting.

We adopt the HAEs' $M_\star$ as a proxy for the stellar masses of AGN host galaxies and present the $M_\star-M_{\rm BH}$ relation of AGNs in Figure \ref{fig:Mbh_Ms_relation}. We adopt the $M_{\rm BH}$ values for the grism-selected broad-line AGNs as representatives (Table \ref{tab:broadline_agn}), which range from $10^{6.85} M_\odot$ to $10^{7.99} M_\odot$ at $3.9 < z < 5$ with a median of $10^{7.32} M_\odot$, and from $10^{6.99} M_\odot$ to $10^{8.29} M_\odot$ at $5 < z < 6$ with a median of $10^{7.46} M_\odot$. Our results are consistent with some of the measurements for individual AGNs \citep{Maiolino2023, Hainline2024}, although the latter exhibit a wide range of $M_\star$ with large uncertainties. On the $M_{\rm BH}-M_\star$ diagram,  these high-redshift AGNs have smaller $M_\star$ compared to those of local AGNs with similar $M_{\rm BH}$. The observed high $M_{\rm BH}/M_\star$ ratio cannot be fully attributed to observational bias \citep{Li2025, SunYang2025}. The $M_{\rm BH}-M_\star$ relation for these AGNs is also offset from the theoretical prediction \citep{Inayoshi2024}, which was developed to model both high-luminosity quasars and low-luminosity AGNs in the early Universe. These overmassive BHs, with higher $M_{\rm BH}/M_\star$ ratios than predicted values, highlight the need for next-generation models to better understand BH seeding mechanisms and the co-evolution of BHs and galaxies.

We also caution that the estimate of \MBH\ is highly uncertain and debated. \citet{Rusakov2025} suggested that Balmer lines may be broadened by electron scattering in regions with high electron column densities. This effect could lead to an overestimation of \MBH\ by 1–2 dex. In contrast, \citet{Juodvzbalis2025} argued that this claim is untenable. Regardless, the estimates of $M_h$ and the derived $M_\star$ from $M_h$ are independent of \MBH. If \MBH\ is overestimated, and the true mass is $10^5$–$10^7\ M_\odot$ as suggested by \citet{Rusakov2025}, the \MBH-$M_\star$ relation would better match the model prediction.

\bigskip

\section{Summary}\label{sec:summary}

In this paper, we study the large-scale environments of low-luminosity AGNs at $3.9 < z < 6$ in the GOODS-North field. By combining the JWST/NIRCam F356W grism from the CONGRESS program and the F444W grism from the FRESCO program, we identify \NHAE\ HAEs within this redshift range. From this dataset, we construct a sample of  \NAGN\ low-luminosity AGNs selected using both the grism and NIRSpec spectra, along with spec-$z$-confirmed V-shaped SEDs. Our main conclusions are as follows:

\begin{itemize}
    \item Low-luminosity AGNs reside in a diverse range of large-scale environments. These AGNs are found in regions with overdensity fields $\delta$  within 15 cMpc spanning from low densities of $\delta=-0.56$ to high densities of $\delta$=10.56. Notably, five AGNs are located in a protocluster at $z = 4.41$ and seven AGNs lie in filamentary structures at $z \approx 5.19$. Among the \NAGN\ AGNs, 10 ($36\pm9$\%) are located in regions with overdensity $\delta > 3$, a fraction consistent with that of HAEs in $\delta > 3$ regions ($30\pm2$\%). This implies that AGNs do not preferentially reside in denser environments compared to HAEs.
    
    \item No clear correlations are found between the overdensity field ($\delta$; within $15\, h^{-1}$ cMpc of AGNs) and their broad \ha\ luminosities, broad \ha\ FWHMs, or BH masses. The AGN fraction among star-forming galaxies also shows no dependence on $\delta$. These results suggest that large-scale environments ($>10$\,cMpc) may not significantly influence black hole growth and AGN evolution on smaller (pc) scales.

    \item The projected surface density excess of AGNs is consistent with that of HAEs, indicating that star-forming galaxies do not cluster more strongly around AGNs. This qualitatively indicates that the dark matter halos of AGNs have masses similar to those of star-forming galaxies.

    \item The cross-correlations between AGNs and HAEs exhibit an amplitude comparable to that of the HAE auto-correlations at both $3.9 < z < 5$ and $5 < z < 6$. When considering AGNs and HAEs in the fields alone, the correlation length of the AGN-HAE cross-correlation is $4.26^{+0.70}_{-0.61}$ $h^{-1}$cMpc with a power-law index $\gamma = 1.59$ at $3.9 < z < 5$, and $7.66^{+0.90}_{-0.81}$ $h^{-1}$cMpc with $\gamma = 1.63$ at $5 < z < 6$. These values are consistent with those of the HAE auto-correlations within $1\sigma$. It suggests that AGNs and HAEs share similar bias parameters ($b_a \approx b_g$) and, consequently, reside in dark matter halos of comparable mass.

    \item  Adopting the halo mass distribution for HAEs derived using the \textsc{UniverseMachine} simulation \citep{Lin_GDN_HAE}, we find that low-luminosity AGNs are hosted by dark matter halos with masses of $\log(M_h/M_\odot) = 11.0-11.2$, with $1\sigma$ scattering of 0.3–0.4 dex. Their $M_h$ values are 1-2 dex lower than those of luminous quasars at similar redshifts. The less biased host dark matter halos suggest that low-luminosity AGNs likely represent a distinct evolutionary phase or AGN population. Interestingly, our $M_h \approx 10^{11} M_\odot$ estimate is consistent with the low-duty cycle scenarios required for super-Eddington accretion as suggested in simulations \citep{Pizzati2024}.

    \item Assuming the same stellar-to-halo mass ratio for AGNs and HAEs, the stellar masses ($M_\star$) of AGN host galaxies are $\log(M_\star/M_\odot) = 8.2-8.4$  with a typical $1\sigma$ scattering of 0.8 dex. The low-luminosity AGNs have overmassive BHs, showing higher $M_\star$/$M_{\rm BH}$ ratios compared to local type-1 AGNs and theoretical predictions \citep{Inayoshi2024}.   

\end{itemize}

To better understand the nature of the JWST-selected low-luminosity AGNs, deep, wider spectroscopic surveys with large AGN samples across a wide range of BH masses and luminosity ranges are required. Deep spectroscopic surveys with large AGN samples, such as SAPPHIRES \citep[GO 6434, PI Egami,][]{sapphires_edr},  will bring insights into the impact of local conditions on BH growth by small-scale clustering analysis. Wide grism surveys, like COSMOS-3D (GO 5893; PI Kakiichi) and NEXUS (GO 5105; PI Shen, \citealt{Shen2024}), will provide better constraints on the halo masses across different luminosities, while also helping to mitigate cosmic variance.

\bigskip

\section*{Acknowledgments}

We thank the anonymous referee for providing constructive comments. We thank Nickolas Kokron, Michael Strauss, and Yin Li for very helpful discussions on the clustering analysis. X.L. and X.F. acknowledge support from the NSF award AST-2308258. F.W. acknowledges support from  NSF award AST-2513040. X.L. and Z.C. acknowledge support from the National Key R\&D Program of China (grant no. 2023YFA1605600) and Tsinghua University Initiative Scientific Research Program (No. 20223080023). A.J.B acknowledges funding from the “FirstGalaxies” Advanced Grant from the European Research Council (ERC) under the European Union’s Horizon 2020 research and innovation program (Grant agreement No. 789056). B.E.R acknowledges support from the NIRCam Science Team contract to the University of Arizona, NAS5-02015, and JWST Program 3215. ST acknowledges support by the Royal Society Research Grant G125142. C.N.A.W acknowledges JWST/NIRCam contract to the University of Arizona NAS5-02015. R.M. acknowledges support by the Science and Technology Facilities Council (STFC), by the ERC through Advanced Grant 695671 “QUENCH”, and by the UKRI Frontier Research grant RISE and FALL. R.M. also acknowledges funding from a research professorship from the Royal Society.

This work is based on observations made with the NASA/ESA Hubble Space Telescope and NASA/ESA/CSA James Webb Space Telescope. The data were obtained from the Mikulski Archive for Space Telescopes at the Space Telescope Science Institute, which is operated by the Association of Universities for Research in Astronomy, Inc., under NASA contract NAS 5-03127 for JWST. These observations are associated with program Nos. 1181 (JADES), 1895 (FRESCO), and 3577 (CONGRESS). 
Support for program No. 3577 was provided by NASA through a grant from the Space Telescope Science Institute, which is operated by the Association of Universities for Research in Astronomy, Inc., under NASA contract NAS 5-03127.
The authors acknowledge the FRESCO team for developing their observing program with a zero-exclusive-access period.

\bigskip
\section*{Data Availability}
The JWST data presented in this article were obtained from the Mikulski Archive for Space Telescopes (MAST) at the Space Telescope Science Institute. The data of the FRESCO survey \citep{FRESCO_hlsp} is available at  \dataset[DOI:10.17909/gdyc-7g80]{https://doi.org/10.17909/gdyc-7g80}; the data of the CONGRESS survey is available at  \dataset[DOI:10.17909/6rfk-6s81]{https://doi.org/10.17909/6rfk-6s81}; the data of the JADES survey \citep{JADES_hlsp} is available at \dataset[DOI:10.17909/8tdj-8n28]{https://doi.org/10.17909/8tdj-8n28}.

\appendix
\counterwithin{figure}{section}

\section{AGN sample}

\begin{figure*}[!ht]
    \centering
    \includegraphics[width=1\linewidth]{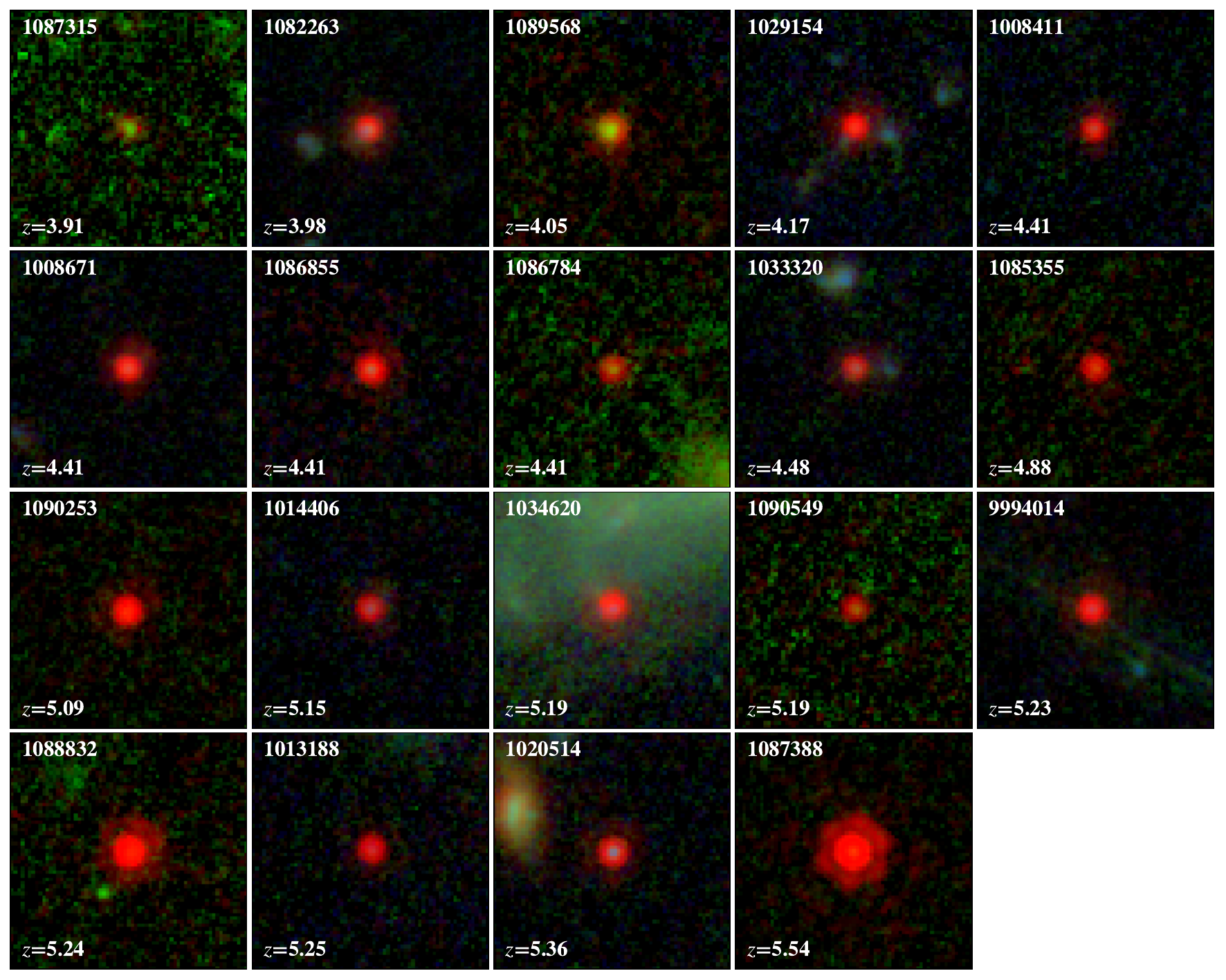}
    \caption{The 19 AGNs in the sample, identified through their broad \ha\ emission lines from JWST/NIRCam WFSS. For each AGN, we present a 2\arcsec$\times$2\arcsec thumbnail composed of F356W (or F444W for $z > 5$), F200W, and F115W images. For the sources 1085355, 1086784, 1087315, 1087388, 1088832, 1089568, 1090253, and 1090549, F210M images are used in place of F200W. }
    \label{fig:AGN_sample_WFSS}
\end{figure*}

\begin{figure*}[!ht]
    \centering
    \includegraphics[width=0.8\linewidth]{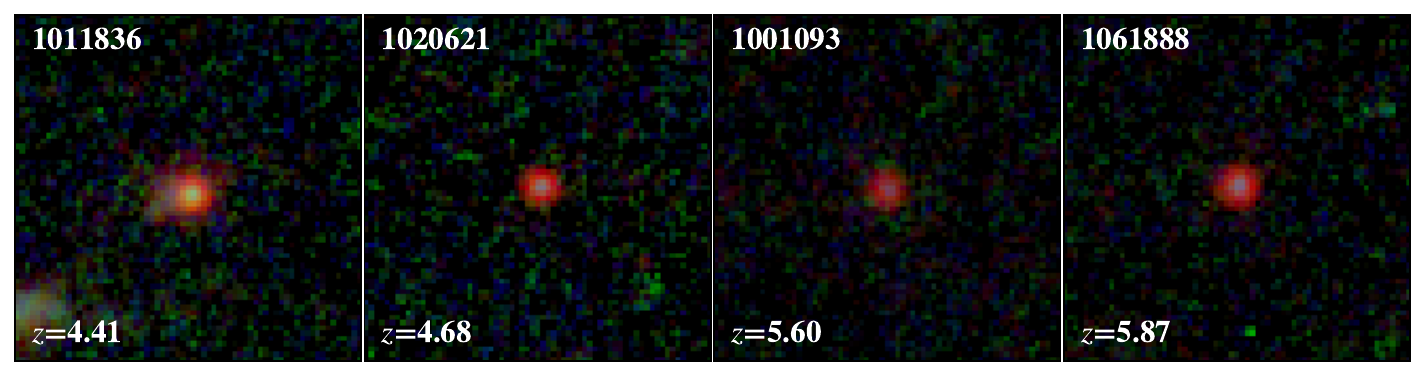}
    \caption{Similar to Figure~\ref{fig:AGN_sample_WFSS} but for the four AGNs in the sample identified through their broad \ha\ emission lines from JWST/NIRSpec Grating R1000 spectra \citep{Maiolino2023}.} 
    \label{fig:AGN_sample_NIRSpec}
\end{figure*}

\begin{figure*}[!ht]
    \centering
    \includegraphics[width=1\linewidth]{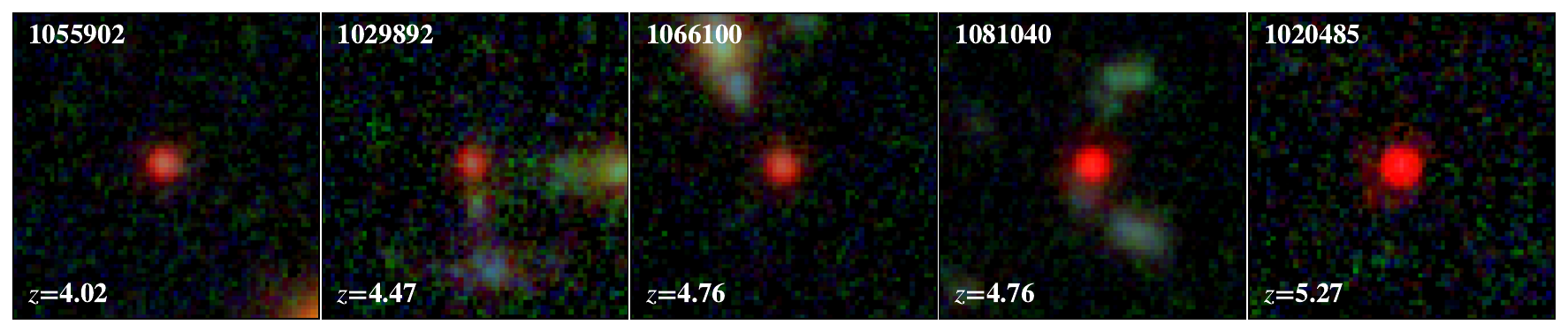}
    \caption{Similar to Figure~\ref{fig:AGN_sample_WFSS} but for the five AGNs in the sample identified through their V-shaped SEDs.}
    \label{fig:AGN_sample_Vshape}
\end{figure*}

\begin{figure*}[!ht]
    \centering
    \includegraphics[width=1\textwidth]{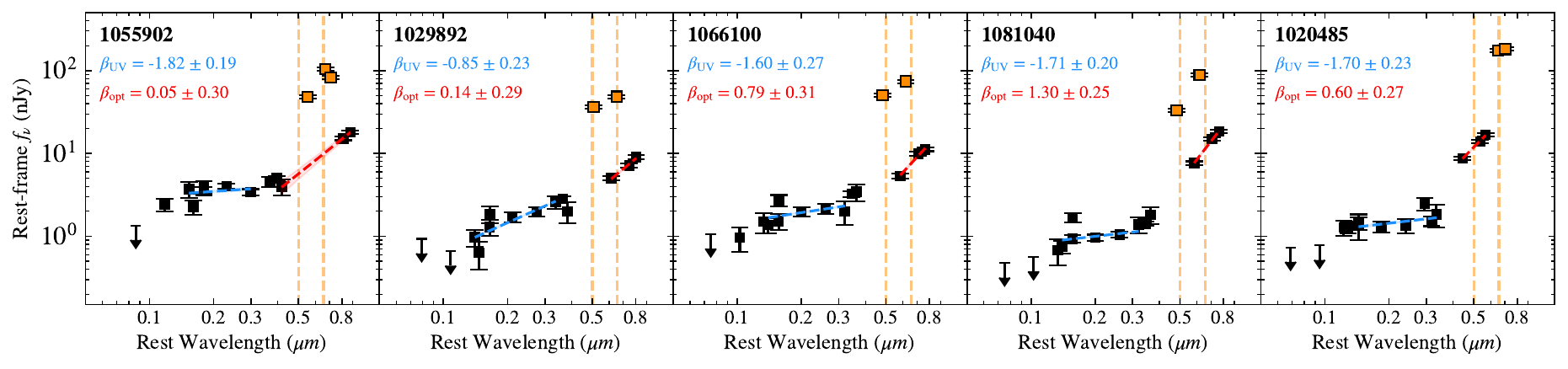}
    \caption{The rest-frame SEDs of the five V-shaped-selected AGNs. All the photometry shown is measured with Kron radii of 1.2. The continuum photometry is represented by black squares, while the bands containing emission lines (H$\beta$, [\ion{O}{3}] $\lambda\lambda$4960,5008, and \ha) are marked as orange squares. The wavelengths of the emission lines are indicated by dashed orange vertical lines. The blue and red dashed lines represent the best-fit power-law models to the UV and optical continuum SEDs, respectively, with the corresponding shaded regions indicating the 1$\sigma$ uncertainty. }
    \label{fig:VshapeSED}
\end{figure*}

We show the NIRCam images of our grism-selected AGN sample in Figure \ref{fig:AGN_sample_WFSS}, NIRSpec-selected sample in Figure \ref{fig:AGN_sample_NIRSpec}, and V-shaped-selected sample in Figure \ref{fig:AGN_sample_Vshape}.  For the V-shaped-selected sample, we present their rest-frame SEDs in Figure \ref{fig:VshapeSED}, which clearly exhibit bluer UV continuum slopes than the optical continuum slopes.

\bigskip

\bibliography{main}{}
\bibliographystyle{aasjournal}

\end{document}